\documentclass[11pt]{article}
\usepackage{amsfonts}
\usepackage{graphicx}
\usepackage{amsmath}
\usepackage{amssymb}
\usepackage{caption2}
\begin{document}
\bibliographystyle{prsty}
\begin{center}
{\large {\bf \sc{  Mass spectrum of the vector hidden
charm and bottom tetraquark states }}} \\[2mm]
Zhi-Gang Wang \footnote{E-mail,wangzgyiti@yahoo.com.cn.  }     \\
 Department of Physics, North China Electric Power University,
Baoding 071003, P. R. China
\end{center}

\begin{abstract}
In this article, we  perform a systematic study of  the mass
spectrum of the vector hidden charm and bottom tetraquark states
using the QCD sum rules.
\end{abstract}



\section{Introduction}

The Babar, Belle, CLEO, D0, CDF and FOCUS collaborations have
discovered (or confirmed) a large number of charmonium-like states,
such as  $X(3940)$, $X(3872)$, $Y(4260)$, $Y(4008)$, $Y(3940)$,
$Y(4325)$, $Y(4360)$, $Y(4660)$, etc,  and revitalized  the interest
in the spectroscopy of the charmonium states
\cite{review1,review2,review3,review4,Olsen2009}. Many possible
assignments for those states have been suggested, such as multiquark
states (irrespective of the molecule  type and the
diquark-antidiquark type), hybrid states, charmonium states modified
by nearby thresholds, threshold cusps, etc
\cite{review1,review2,review3,review4}.

  The $Z^+(4430)$  observed in the  decay mode $\psi^\prime\pi^+$ by the
 Belle collaboration is the most
interesting subject \cite{Belle-z4430}.  We can distinguish the
multiquark states
 from the hybrids or charmonia with the criterion of
non-zero charge. The $Z^+(4430)$ can't be a pure $c\bar{c}$ state
due to the positive charge,   and  may be a $c\bar{c}u\bar{d}$
tetraquark state. However, the Babar collaboration did not confirm
this resonance \cite{Babar0811}. The two resonance-like structures
 $Z(4050)$ and $Z(4250)$
 in the $\pi^+\chi_{c1}$ invariant mass distribution
near $4.1 \,\rm{GeV}$ are also particularly  interesting
\cite{Belle-chipi}. Their quark contents must be some special
combinations of the $c\bar{c} u\bar{d}$, just like the $Z^+(4430)$,
they can't be the conventional mesons.

In Refs.\cite{Wang0807,Wang08072}, we assume  that the hidden charm
mesons $Z(4050)$ and $Z(4250)$ are vector (and scalar) tetraquark
states, and study their masses using the QCD sum rules. The
numerical results indicate that the mass of the vector hidden charm
tetraquark state is about $M_{Z}=(5.12\pm0.15)\,\rm{GeV}$ or
$M_{Z}=(5.16\pm0.16)\,\rm{GeV}$, while the mass of the scalar hidden
charm tetraquark state
 is about $M_{Z}=(4.36\pm0.18)\,\rm{GeV}$. The resonance-like structure  $Z(4250)$ observed by the Belle
collaboration  in the $\pi^+\chi_{c1}$ invariant mass distribution
near $4.1 \,\rm{GeV}$ in the exclusive decays $\bar{B}^0\to K^-
\pi^+ \chi_{c1}$  can be tentatively identified as the scalar
tetraquark state \cite{Wang08072}. In Ref.\cite{Wang0902}, we study
the mass spectrum of the scalar hidden charm and bottom tetraquark
states using the QCD sum rules. In this article, we extend our
previous work to study the mass spectrum of the vector hidden charm
and bottom tetraquark states.

In the QCD sum rules, the operator product expansion is used to
expand the time-ordered currents into a series of quark and gluon
condensates which parameterize the long distance properties of  the
QCD vacuum. Based on the quark-hadron duality, we can obtain copious
information about the hadronic parameters at the phenomenological
side \cite{SVZ79,Reinders85}.

The mass is a fundamental parameter in describing a hadron, whether
or not there exist those hidden charm or bottom tetraquark
configurations is of great importance itself, because it provides a
new opportunity for a deeper understanding of the low energy QCD.
The  vector hidden charm ($c\bar{c}$) and bottom ($b\bar{b}$)
tetraquark states may be observed at the LHCb, where the $b\bar{b}$
pairs will be copiously produced with the cross section about $500
\,\mu b$ \cite{LHC}.

The hidden charm and bottom tetraquark states ($Z$) have the
symbolic quark structures:
\begin{align}
  Z^+ = Q\bar{Q} u  \bar{d}  ;~~~~
  Z^0 = \frac{1}{\sqrt{2}}Q\bar{Q}&( u  \bar{u}-d  \bar{d})  ;~~~~
  Z^- =Q\bar{Q}d\bar{u}    ; \nonumber\\
  Z_s^+ = Q\bar{Q}u  \bar{s} ;~~~~
  Z_s^- =Q\bar{Q}  s\bar{u}  ;&~~~~
  Z_s^0 = Q\bar{Q}d  \bar{s} ;~~~~
  \overline Z_s^0 = Q\bar{Q}s\bar{d} ; \nonumber \\
  Z_\varphi= \frac{1}{\sqrt{2}} Q\bar{Q} (u\bar{u}+d\bar{d});
  &~~~~Z_\phi =  Q\bar{Q} s\bar{ s} \, ,
\end{align}
where the $Q$ denote the heavy quarks $c$ and $b$.

We take the diquarks as the basic constituents   following  Jaffe
and Wilczek \cite{Jaffe2003,Jaffe2004}, and construct the tetraquark
states with the diquark and antidiquark pairs.  The diquarks have
five Dirac tensor structures, scalar $C\gamma_5$, pseudoscalar $C$,
vector $C\gamma_\mu \gamma_5$, axial vector $C\gamma_\mu $  and
tensor $C\sigma_{\mu\nu}$, where  $C$ is the charge conjunction
matrix. The structures $C\gamma_\mu $ and $C\sigma_{\mu\nu}$ are
symmetric, the structures $C\gamma_5$, $C$ and $C\gamma_\mu
\gamma_5$ are antisymmetric. The attractive interactions of
one-gluon exchange favor  formation of the diquarks in  color
antitriplet $\overline{3}_{ c}$, flavor antitriplet $\overline{3}_{
f}$ and spin singlet $1_s$ \cite{GI1,GI2}.   In this article, we
assume the vector hidden charm and bottom  mesons $Z$ consist of the
$C\gamma_5-C\gamma_\mu \gamma_5$ type and $C-C\gamma_\mu $ type
diquark structures, and construct the interpolating currents
$J^\mu(x)$ and $\eta^\mu(x)$:
\begin{eqnarray}
J^\mu_{Z^+}(x)&=& \epsilon^{ijk}\epsilon^{imn}u_j^T(x) C\gamma_5
Q_k(x)\bar{Q}_m(x) \gamma_5 \gamma^\mu C \bar{d}_n^T(x)\, , \nonumber\\
J^\mu_{Z^0}(x)&=& \frac{\epsilon^{ijk}\epsilon^{imn}}{\sqrt{2}}
\left[u_j^T(x) C\gamma_5 Q_k(x) \bar{Q}_m(x) \gamma_5 \gamma^\mu C
\bar{u}_n^T(x)-(u\rightarrow d)\right]\, , \nonumber\\
J^\mu_{Z^+_s}(x)&=& \epsilon^{ijk}\epsilon^{imn}u_j^T(x) C\gamma_5
Q_k(x)\bar{Q}_m(x) \gamma_5 \gamma^\mu C \bar{s}_n^T(x)\, , \nonumber\\
J^\mu_{Z^0_s}(x)&=& \epsilon^{ijk}\epsilon^{imn}d_j^T(x) C\gamma_5
Q_k(x)\bar{Q}_m(x) \gamma_5 \gamma^\mu C \bar{s}_n^T(x)\, , \nonumber\\
J^\mu_{Z_\varphi}(x)&=&
\frac{\epsilon^{ijk}\epsilon^{imn}}{\sqrt{2}} \left[u_j^T(x)
C\gamma_5 Q_k(x) \bar{Q}_m(x) \gamma_5 \gamma^\mu C
\bar{u}_n^T(x) +(u\rightarrow d)\right]\, , \nonumber\\
 J^\mu_{Z_\phi}(x)&=& \epsilon^{ijk}\epsilon^{imn}s_j^T(x) C\gamma_5
Q_k(x)\bar{Q}_m(x) \gamma_5 \gamma^\mu C \bar{s}_n^T(x)\, , \nonumber\\
\eta^\mu_{Z^+}(x)&=& \epsilon^{ijk}\epsilon^{imn}u_j^T(x) C
Q_k(x)\bar{Q}_m(x)  \gamma^\mu C \bar{d}_n^T(x)\, , \nonumber\\
\eta^\mu_{Z^0}(x)&=& \frac{\epsilon^{ijk}\epsilon^{imn}}{\sqrt{2}}
\left[u_j^T(x) C Q_k(x) \bar{Q}_m(x)  \gamma^\mu C
\bar{u}_n^T(x)-(u\rightarrow d)\right]\, , \nonumber\\
\eta^\mu_{Z^+_s}(x)&=& \epsilon^{ijk}\epsilon^{imn}u_j^T(x) C
Q_k(x)\bar{Q}_m(x)  \gamma^\mu C \bar{s}_n^T(x)\, , \nonumber\\
\eta^\mu_{Z^0_s}(x)&=& \epsilon^{ijk}\epsilon^{imn}d_j^T(x) C
Q_k(x)\bar{Q}_m(x)  \gamma^\mu C \bar{s}_n^T(x)\, , \nonumber\\
\eta^\mu_{Z_\varphi}(x)&=&
\frac{\epsilon^{ijk}\epsilon^{imn}}{\sqrt{2}} \left[u_j^T(x) C
Q_k(x) \bar{Q}_m(x)  \gamma^\mu C
\bar{u}_n^T(x) +(u\rightarrow d)\right]\, , \nonumber\\
 \eta^\mu_{Z_\phi}(x)&=& \epsilon^{ijk}\epsilon^{imn}s_j^T(x) C
Q_k(x)\bar{Q}_m(x)  \gamma^\mu C \bar{s}_n^T(x)\, ,
\end{eqnarray}
where the $i$, $j$, $k$, $\cdots$  are color indexes. In the isospin
limit, the interpolating currents result in six distinct expressions
for the spectral densities (see Eq.(8)), which are  characterized by
the Dirac structures of the interpolating currents and the number of
the $s$ quark they contain.

We can also interpolate the vector tetraquark states with the
currents $\hat{J}^\mu(x)$ and $\hat{\eta}^\mu(x)$, which consist of
$C\gamma_\mu \gamma_5 -C\gamma_5$ type and $C\gamma_\mu-C$ type
diquark structures, respectively:
\begin{eqnarray}
\hat{J}^\mu_{Z^+}(x)&=& \epsilon^{ijk}\epsilon^{imn}u_j^T(x)
C\gamma^\mu\gamma_5 Q_k(x)\bar{Q}_m(x) \gamma_5  C \bar{d}_n^T(x)\, , \nonumber\\
\hat{J}^\mu_{Z^0}(x)&=&
\frac{\epsilon^{ijk}\epsilon^{imn}}{\sqrt{2}} \left[u_j^T(x)
C\gamma^\mu\gamma_5 Q_k(x) \bar{Q}_m(x) \gamma_5  C
\bar{u}_n^T(x)-(u\rightarrow d)\right]\, , \nonumber\\
\hat{J}^\mu_{Z^+_s}(x)&=& \epsilon^{ijk}\epsilon^{imn}u_j^T(x)
C\gamma^\mu\gamma_5
Q_k(x)\bar{Q}_m(x) \gamma_5  C \bar{s}_n^T(x)\, , \nonumber\\
\hat{J}^\mu_{Z^0_s}(x)&=& \epsilon^{ijk}\epsilon^{imn}d_j^T(x)
C\gamma^\mu\gamma_5Q_k(x)\bar{Q}_m(x) \gamma_5  C \bar{s}_n^T(x)\, , \nonumber\\
\hat{J}^\mu_{Z_\varphi}(x)&=&
\frac{\epsilon^{ijk}\epsilon^{imn}}{\sqrt{2}} \left[u_j^T(x)
C\gamma^\mu\gamma_5 Q_k(x) \bar{Q}_m(x) \gamma_5  C
\bar{u}_n^T(x) +(u\rightarrow d)\right]\, , \nonumber\\
 \hat{J}^\mu_{Z_\phi}(x)&=& \epsilon^{ijk}\epsilon^{imn}s_j^T(x) C\gamma^\mu\gamma_5
Q_k(x)\bar{Q}_m(x) \gamma_5  C \bar{s}_n^T(x)\, , \nonumber\\
\hat{\eta}^\mu_{Z^+}(x)&=& \epsilon^{ijk}\epsilon^{imn}u_j^T(x)
C\gamma^\mu Q_k(x)\bar{Q}_m(x)   C \bar{d}_n^T(x)\, , \nonumber\\
\hat{\eta}^\mu_{Z^0}(x)&=&
\frac{\epsilon^{ijk}\epsilon^{imn}}{\sqrt{2}} \left[u_j^T(x)
C\gamma^\mu Q_k(x) \bar{Q}_m(x)   C
\bar{u}_n^T(x)-(u\rightarrow d)\right]\, , \nonumber\\
\hat{\eta}^\mu_{Z^+_s}(x)&=& \epsilon^{ijk}\epsilon^{imn}u_j^T(x)
C\gamma^\mu Q_k(x)\bar{Q}_m(x)   C \bar{s}_n^T(x)\, , \nonumber\\
\hat{\eta}^\mu_{Z^0_s}(x)&=& \epsilon^{ijk}\epsilon^{imn}d_j^T(x)
C\gamma_\mu Q_k(x)\bar{Q}_m(x)  C \bar{s}_n^T(x)\, , \nonumber\\
\hat{\eta}^\mu_{Z_\varphi}(x)&=&
\frac{\epsilon^{ijk}\epsilon^{imn}}{\sqrt{2}} \left[u_j^T(x)
C\gamma_\mu Q_k(x) \bar{Q}_m(x)  C
\bar{u}_n^T(x) +(u\rightarrow d)\right]\, , \nonumber\\
\hat{\eta}^\mu_{Z_\phi}(x)&=& \epsilon^{ijk}\epsilon^{imn}s_j^T(x)
C\gamma_\mu Q_k(x)\bar{Q}_m(x)  C \bar{s}_n^T(x)\, .
\end{eqnarray}
 Our analytical results indicate that
the interpolating currents $J^\mu(x)$ ($\eta^\mu(x)$) and
$\hat{J}^\mu(x)$ ($\hat{\eta}^\mu(x)$) lead to the same expression
for the correlation functions $\Pi_{\mu\nu}(p)$, for example,
\begin{align}
  J^\mu_{Z^+} \sim  \hat{J}^\mu_{Z^+}  ;~~~~
  J^\mu_{Z^0} \sim & \hat{J}^\mu_{Z^0}  ;~~~~
  J^\mu_{Z^-} \sim  \hat{J}^\mu_{Z^-}    ; \nonumber\\
  J^\mu_{Z_s^+} \sim \hat{J}^\mu_{Z_s^+} ;~~~~
  J^\mu_{Z_s^-} \sim \hat{J}^\mu_{Z_s^-}  ;&~~~~
  J^\mu_{Z_s^0} \sim \hat{J}^\mu_{Z_s^0} ;~~~~
  J^\mu_{\bar{Z}_s^+} \sim \hat{J}^\mu_{\bar{Z}_s^+} ; \nonumber \\
  J^\mu_{Z_\varphi}\sim \hat{J}^\mu_{Z_\varphi};
  &~~~~J^\mu_{Z_\phi}\sim \hat{J}^\mu_{Z_\phi} \, ,
\end{align}
where we use $\sim$ to denote the two interpolating  currents lead
to the same expression.   The special superpositions
$tJ^\mu(x)+(1-t)\hat{J}^\mu(x)$ and
$t\eta^\mu(x)+(1-t)\hat{\eta}^\mu(x)$  can't  improve the
predictions remarkably, where $t=0-1$. In this article, we take only
the interpolating currents $J^\mu(x)$ and $\eta^\mu(x)$ for
simplicity, i.e. $t=1$; the explicit expressions of the
corresponding spectral densities are shown in Eq.(8) and
Eqs.(10-12).

The article is arranged as follows:  we derive the QCD sum rules for
  the vector hidden charm and bottom tetraquark states $Z$  in section 2; in section 3, numerical
results and discussions; section 4 is reserved for conclusion.

\section{QCD sum rules for  the vector  tetraquark states $Z$ }
In the following, we write down  the two-point correlation functions
$\Pi_{\mu\nu}(p)$  in the QCD sum rules,
\begin{eqnarray}
\Pi_{\mu\nu}(p)&=&i\int d^4x e^{ip \cdot x} \langle
0|T\left[J/\eta_\mu(x)J/\eta_\nu^{\dagger}(0)\right]|0\rangle \, ,
\end{eqnarray}
where the  $J^\mu(x)$ ($\eta^\mu(x)$) denotes the interpolating
currents $J^\mu_{Z^+}(x)$ ($\eta^\mu_{Z^+}(x)$), $J^\mu_{Z^0}(x)$
($\eta^\mu_{Z^0}(x)$), $J^\mu_{Z^+_s}(x)$ ($\eta^\mu_{Z^+_s}(x)$),
etc.

We can insert  a complete set of intermediate hadronic states with
the same quantum numbers as the current operators $J_\mu(x)$ and
$\eta_\mu(x)$ into the correlation functions  $\Pi_{\mu\nu}(p)$  to
obtain the hadronic representation \cite{SVZ79,Reinders85}. After
isolating the ground state contribution from the pole term of the
$Z$, we get the following result,
\begin{eqnarray}
\Pi_{\mu\nu}(p)&=&\frac{\lambda_{Z}^2}{M_{Z}^2-p^2}\left[-g_{\mu\nu}+\frac{p_\mu
p_\nu}{p^2} \right] +\cdots \, \, ,
\end{eqnarray}
where the pole residue (or coupling) $\lambda_Z$ is defined by
\begin{eqnarray}
\lambda_{Z} \epsilon_\mu  &=& \langle 0|J/\eta_\mu(0)|Z(p)\rangle \,
,
\end{eqnarray}
 the $\epsilon_\mu$ denotes the polarization vector.

 After performing the standard procedure of the QCD sum rules, we obtain the following  twelve   sum rules:
\begin{eqnarray}
\lambda_{\pm i}^2 e^{-\frac{M_{\pm i}^2}{M^2}}= \int_{\Delta_{\pm
i}}^{s^0_{\pm i}} ds \rho_i^{\pm}(s)e^{-\frac{s}{M^2}} \, ,
\end{eqnarray}
where the $i$ denote the $c\bar{c}q\bar{q}$,
   $c\bar{c}q\bar{s}$, $c\bar{c}s\bar{s}$, $b\bar{b}q\bar{q}$,
   $b\bar{b}q\bar{s}$ and $b\bar{b}s\bar{s}$ channels, respectively;
   the $s_i^0$ are the corresponding continuum threshold parameters, the $\pm$ denote the current operators of the
  $C\gamma_5-C\gamma_\mu \gamma_5$ type and $C-C\gamma_\mu $ type respectively;  and the $M^2$ is the Borel
  parameter. The thresholds $\Delta_{\pm
i}$ can be sorted into three sets,  we introduce the $q\bar{q}$,
$q\bar{s}$ and $s\bar{s}$ to denote the light quark constituents in
the vector tetraquark states to simplify the notation,
$\Delta_{q\bar{q}}=4m_Q^2$, $\Delta_{q\bar{s}}=(2m_Q+m_s)^2$,
$\Delta_{s\bar{s}}=4(m_Q+m_s)^2$.
   The explicit expressions of the  spectral densities $\rho^{\pm}_{q\bar{q}}(s)$,
$\rho^{\pm}_{q\bar{s}}(s)$ and $\rho^{\pm}_{s\bar{s}}(s)$ are
presented in the appendix,  where
$\alpha_{max}=\frac{1+\sqrt{1-4m_Q^2/s}}{2}$,
$\alpha_{min}=\frac{1-\sqrt{1-4m_Q^2/s}}{2}$,
$\beta_{min}=\frac{\alpha m_Q^2}{\alpha s -m_Q^2}$,
$\widetilde{m}_Q^2=\frac{(\alpha+\beta)m_Q^2}{\alpha\beta}$,
$\widetilde{\widetilde{m}}_Q^2=\frac{m_Q^2}{\alpha(1-\alpha)}$.

 We carry out the operator
product expansion to the vacuum condensates adding up to
dimension-10 and take analogous  assumptions as in the QCD  sum
rules for the H-dibaryon \cite{Hbaryon}.

$\bullet$ In calculation, we
 take   vacuum saturation for the high
dimension vacuum condensates, they  are always
 factorized to lower condensates with vacuum saturation in the QCD sum rules,
  factorization works well in  large $N_c$ limit. In reality, $N_c=3$, some  ambiguities may come from
the vacuum saturation assumption.

$\bullet$ We take into account the contributions from the quark
condensates,  mixed condensates, and neglect the contributions from
the gluon condensate. The gluon condensate $\langle
\frac{\alpha_sGG}{\pi}\rangle$  is of higher order in $\alpha_s$,
and its contributions   are suppressed by  very  large denominators
comparing with the four quark condensate $\langle \bar{q}q\rangle^2$
(or $\langle \bar{s}s\rangle^2$).  One can consult the sum rules for
the light tetraquark states \cite{Wang1,Wang2}, the heavy tetraquark
state \cite{Wang08072} and the  heavy molecular state
\cite{Wang0904} for example. The gluon condensate $\langle
\frac{\alpha_sGG}{\pi}\rangle$ would not play any significant role,
although the gluon  condensate $\langle
\frac{\alpha_sGG}{\pi}\rangle$ has smaller dimension of mass than
the four quark condensate $\langle \bar{q}q\rangle^2$ (or $\langle
\bar{s}s\rangle^2$). Furthermore, there are many terms involving the
gluon condensate for the heavy tetraquark states and heavy molecular
states in the operator product expansion (one can consult
Refs.\cite{Wang08072,Wang0904}), we neglect the gluon condensate for
simplicity.

$\bullet$ We neglect the terms proportional to the $m_u$ and $m_d$,
their contributions are of minor importance due to the small values
of the $u$ and $d$ quark masses.

 Differentiating  the Eq.(8) with respect to  $\frac{1}{M^2}$, then eliminate the
 pole residues $\lambda_{\pm i}$, we can obtain the sum rules for
 the masses  of the $Z$,
 \begin{eqnarray}
 M_{\pm i}^2= \frac{\int_{\Delta_{\pm i}}^{s^0_{\pm i}} ds \frac{d}{d(-1/M^2)}
\rho^{\pm}_i(s)e^{-\frac{s}{M^2}} }{\int_{\Delta_{\pm i}}^{s^0_{\pm
i}} ds \rho^{\pm}_i(s)e^{-\frac{s}{M^2}}}\, .
\end{eqnarray}

\section{Numerical results and discussions}
The input parameters are taken to be the standard values $\langle
\bar{q}q \rangle=-(0.24\pm 0.01 \,\rm{GeV})^3$, $\langle \bar{s}s
\rangle=(0.8\pm 0.2 )\langle \bar{q}q \rangle$, $\langle
\bar{q}g_s\sigma Gq \rangle=m_0^2\langle \bar{q}q \rangle$, $\langle
\bar{s}g_s\sigma Gs \rangle=m_0^2\langle \bar{s}s \rangle$,
$m_0^2=(0.8 \pm 0.2)\,\rm{GeV}^2$,  $m_s=(0.14\pm0.01)\,\rm{GeV}$,
$m_c=(1.35\pm0.10)\,\rm{GeV}$ and $m_b=(4.8\pm0.1)\,\rm{GeV}$ at the
energy scale  $\mu=1\, \rm{GeV}$ \cite{SVZ79,Reinders85,Ioffe2005}.

The heavy quark mass appearing in the perturbative terms (see e.g.
$\rho_{s\bar{s}}^\pm(s)$) is usually taken to be the pole mass in
the QCD sum rules, while the choice of the $m_Q$ in the
leading-order coefficients of the higher-dimensional terms (vacuum
condensates) is arbitrary \cite{Kho9801}.  The $\overline{MS}$ mass
$m_Q(m_Q^2)$ relates with the pole mass $\hat{m}_Q$ through the
relation $m_Q(m_Q^2) =\hat{m}_Q\left[1+C_F
\alpha_s(m_Q^2)/\pi+\cdots\right]^{-1}$ \cite{PDG}. In this article,
we can take the approximation
$m_Q(\mu^2=1\,\rm{GeV}^2)\approx\hat{m}_Q$ for all the $m_Q$ without
the $\alpha_s$ corrections for consistency. The vacuum condensates
are scale dependent,  one can also   choose the typical scale
$\mu^2=\mathcal {O}(M^2)$, which characterizes  the average
virtuality of the quarks.   As the physical quantities would not
depend on the special energy scale we choose, we expect that scale
dependence of the input parameters is canceled out approximately
with each other, the  masses of the vector tetraquark states  which
are calculated at the energy scale $\mu=1\rm{GeV}$ can make robust
predictions; furthermore, at the energy scale $\mu=1\rm{GeV}$,
perturbative calculations are reliable.

In the conventional QCD sum rules \cite{SVZ79,Reinders85}, there are
two criteria (pole dominance and convergence of the operator product
expansion) for choosing  the Borel parameter $M^2$ and threshold
parameter $s_0$. The light tetraquark states can not satisfy the two
criteria, although it is not an indication of non-existence of the
light tetraquark states (for detailed discussions about this
subject, one can consult Refs.\cite{Wang08072,Wang0708}). We impose
the two criteria on the heavy tetraquark states to choose the Borel
parameter $M^2$ and threshold parameter $s_0$.

The  meson  $Z(4250)$   can be tentatively identified as   a scalar
tetraquark state ($c\bar{c}u\bar{d}$), the decay $ Z(4250) \to
\pi^+\chi_{c1}$ can take place with the Okubo-Zweig-Iizuka (OZI)
super-allowed "fall-apart" mechanism, which can take into account
the large total width naturally \cite{Wang08072}.   While the
$Z(4050)$ is difficult to be identified as the  scalar tetraquark
state ($c\bar{c}u\bar{d}$) considering its small mass.  There  still
lack experiential candidates to identify the vector tetraquark
states $c\bar{c}q\bar{q}$, $c\bar{c}q\bar{s}$, $c\bar{c}s\bar{s}$,
$b\bar{b}q\bar{q}$, $b\bar{b}q\bar{s}$ and $b\bar{b}s\bar{s}$.

The contributions from the high dimension vacuum condensates  in the
operator product expansion for the $C\gamma_5-C\gamma_\mu \gamma_5$
type interpolating currents  are shown in Figs.1-2, where (and
thereafter) we
 use the $\langle\bar{q}q\rangle$ to denote the quark condensates
$\langle\bar{q}q\rangle$, $\langle\bar{s}s\rangle$ and the
$\langle\bar{q}g_s \sigma Gq\rangle$ to denote the mixed condensates
$\langle\bar{q}g_s \sigma Gq\rangle$, $\langle\bar{s}g_s \sigma
Gs\rangle$. The contributions from the terms proportional to the
$m_Q$ are less than (or about) $10\%$ at the value $M^2\geq
3.3\,\rm{GeV}^2$ and play minor important roles, we prefer study the
$C\gamma_5-C\gamma_\mu \gamma_5$ type interpolating currents  in
detail   for simplicity, then take the same Borel parameter and
threshold parameter for the corresponding $C -C\gamma_\mu  $ type
interpolating currents.

From the figures, we can see that the contributions from the high
dimension condensates change  quickly with variation of the Borel
parameter at the values $M^2\leq 3.2\,\rm{GeV}^2$ and $M^2\leq
8.5\,\rm{GeV}^2$ for the $c\bar{c}$ channels and $b\bar{b}$ channels
respectively, such an  unstable  behavior  can not lead to sum rules
stable enough, our numerical results confirm this conjecture.  At
the values $M^2\geq 3.4\,\rm{GeV}^2$ and $s_0\geq 30\,\rm{GeV}^2$,
the contributions from the  $\langle \bar{q}q\rangle^2+\langle
\bar{q}q\rangle \langle \bar{q}g_s \sigma Gq\rangle $ term are less
than (or equal) $15\%$ for the $c\bar{c}q\bar{s}$ channel, the
corresponding contributions are even smaller for the
$c\bar{c}q\bar{q}$ and $c\bar{c}s\bar{s}$ channels; the
contributions from the vacuum condensate of the highest dimension
$\langle\bar{q}g_s \sigma Gq\rangle^2$ are less than (or equal)
$1\%$ for all the $c\bar{c}$ channels, we expect the operator
product expansion is convergent in the $c\bar{c}$ channels. At the
values $M^2\geq 8.6\,\rm{GeV}^2$ and $s_0\geq 156\,\rm{GeV}^2$, the
contributions from the  $\langle \bar{q}q\rangle^2+\langle
\bar{q}q\rangle \langle \bar{q}g_s \sigma Gq\rangle $ term are less
than (or equal) $15\%$ for the $b\bar{b}q\bar{s}$ channel, the
corresponding contributions are even smaller for the
$b\bar{b}q\bar{q}$ and $b\bar{b}s\bar{s}$ channels; the
contributions from the vacuum condensate of the highest dimension
$\langle\bar{q}g_s \sigma Gq\rangle^2$ are less than (or equal)
$3\%$ for all the $b\bar{b}$ channels, we expect the operator
product expansion is convergent in the $b\bar{b}$ channels.

In this article, we take the uniform Borel parameter $M^2_{min}$,
i.e. $M^2_{min}\geq 3.4 \, \rm{GeV}^2$ and $M^2_{min}\geq 8.6 \,
\rm{GeV}^2$ for the $c\bar{c}$ channels and $b\bar{b}$ channels,
respectively.

In Fig.3, we show the  contributions from the pole terms with
variation of the Borel parameter and the threshold parameter for the
$C\gamma_5-C\gamma_\mu \gamma_5$ type interpolating currents. The
pole contributions are larger than (or equal) $50\%$ at the value
$M^2 \leq 4.0 \, \rm{GeV}^2 $ and $s_0\geq
30\,\rm{GeV}^2,\,31\,\rm{GeV}^2,\,31\,\rm{GeV}^2$ for the
$c\bar{c}q\bar{q}$,
   $c\bar{c}q\bar{s}$, $c\bar{c}s\bar{s}$
channels respectively, and larger than (or equal) $50\%$ at the
value $M^2 \leq 9.6 \, \rm{GeV}^2 $ and $s_0\geq
156\,\rm{GeV}^2,\,158\,\rm{GeV}^2,\,158\,\rm{GeV}^2$ for  the
$b\bar{b}q\bar{q}$,
   $b\bar{b}q\bar{s}$ and $b\bar{b}s\bar{s}$ channels respectively. Again we
take the uniform Borel parameter $M^2_{max}$, i.e. $M^2_{max}\leq
4.0 \, \rm{GeV}^2$ and $M^2_{max}\leq 9.4 \, \rm{GeV}^2$ (here we
take a slightly smaller $M^2_{max}$ to enhance the pole
contribution) for the $c\bar{c}$ channels and $b\bar{b}$ channels,
respectively.

Based on above discussions, the threshold parameters are taken as
$s_0=(31\pm1)\,\rm{GeV}^2$, $(32\pm1)\,\rm{GeV}^2$,
$(32\pm1)\,\rm{GeV}^2$, $(158\pm2)\,\rm{GeV}^2$,
$(160\pm2)\,\rm{GeV}^2$ and $(160\pm2)\,\rm{GeV}^2$ for the
$c\bar{c}q\bar{q}$,
   $c\bar{c}q\bar{s}$, $c\bar{c}s\bar{s}$, $b\bar{b}q\bar{q}$,
   $b\bar{b}q\bar{s}$ and $b\bar{b}s\bar{s}$ channels, respectively;
   the Borel parameters are taken as $M^2=(3.4-4.0)\,\rm{GeV}^2$ and
   $(8.6-9.4)\,\rm{GeV}^2$ for the
$c\bar{c}$ channels and $b\bar{b}$ channels, respectively.
      In those regions, the two criteria of the QCD sum rules
are full filled \cite{SVZ79,Reinders85}.

\begin{figure}
 \centering
 \includegraphics[totalheight=5cm,width=6cm]{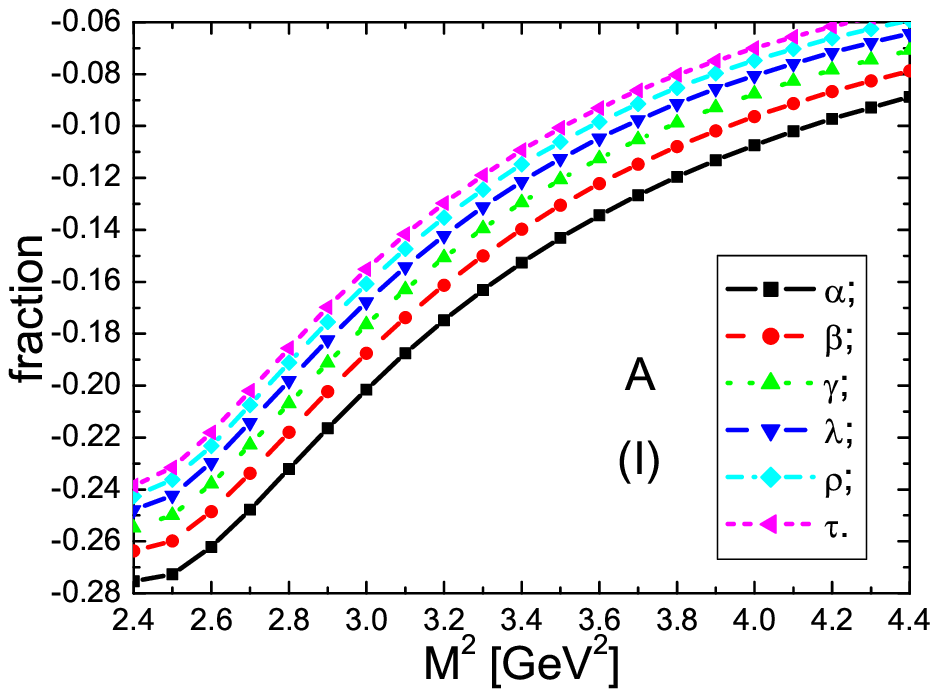}
 \includegraphics[totalheight=5cm,width=6cm]{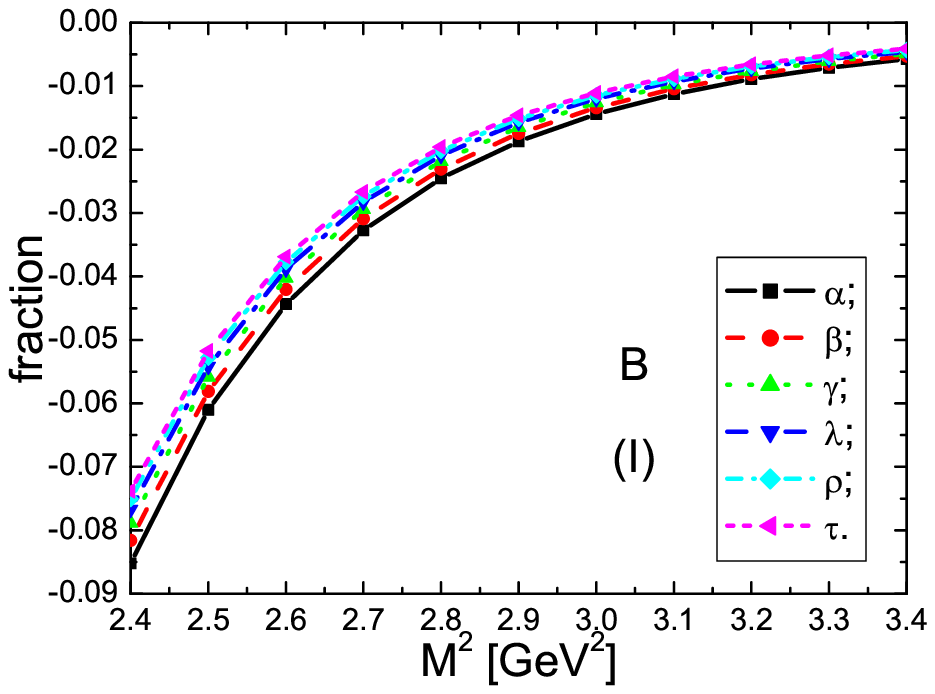}
 \includegraphics[totalheight=5cm,width=6cm]{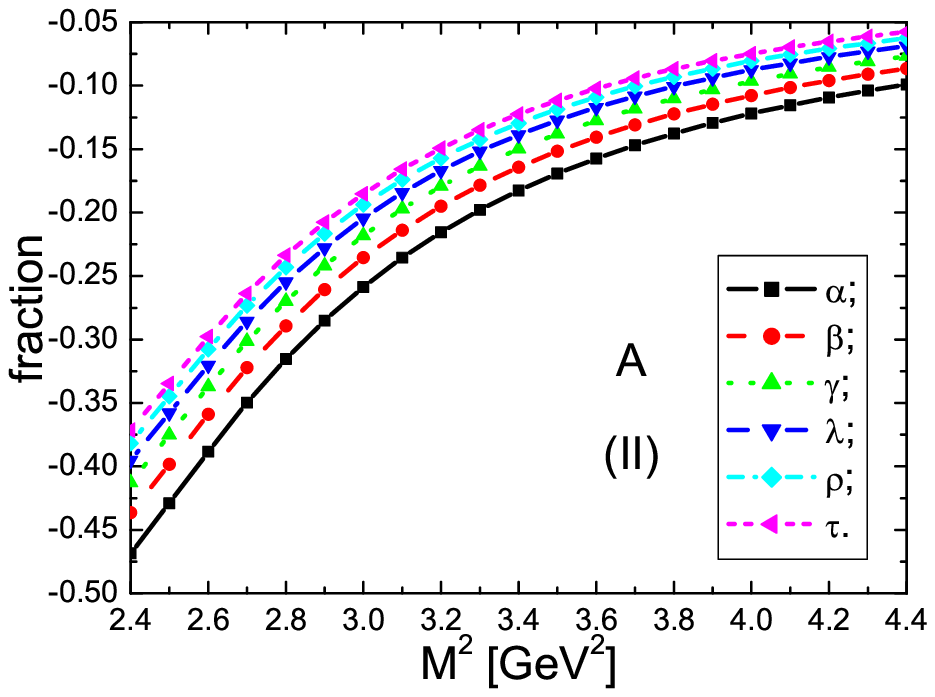}
 \includegraphics[totalheight=5cm,width=6cm]{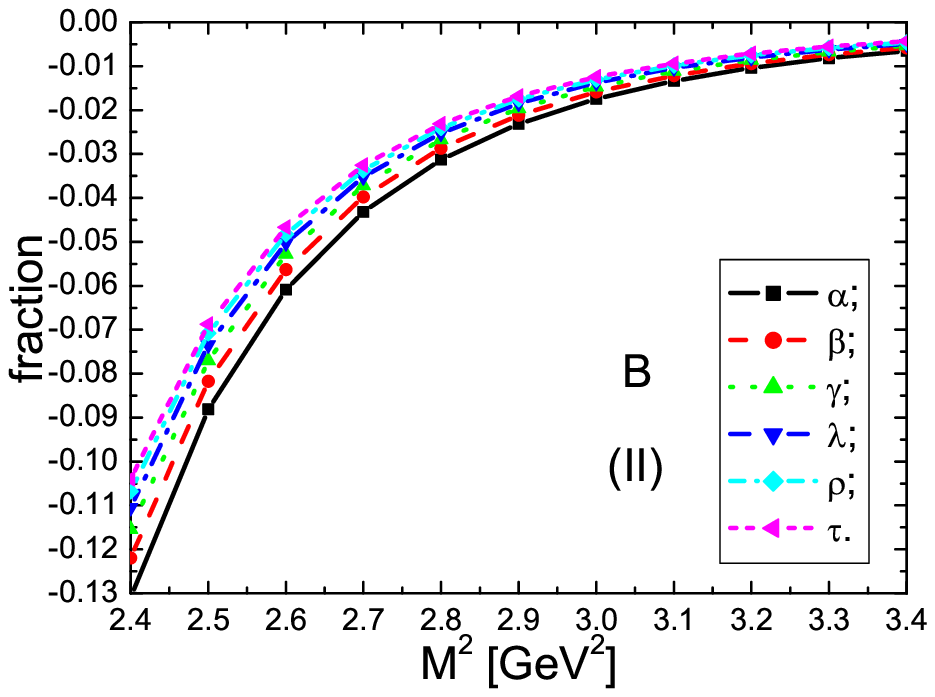}
 \includegraphics[totalheight=5cm,width=6cm]{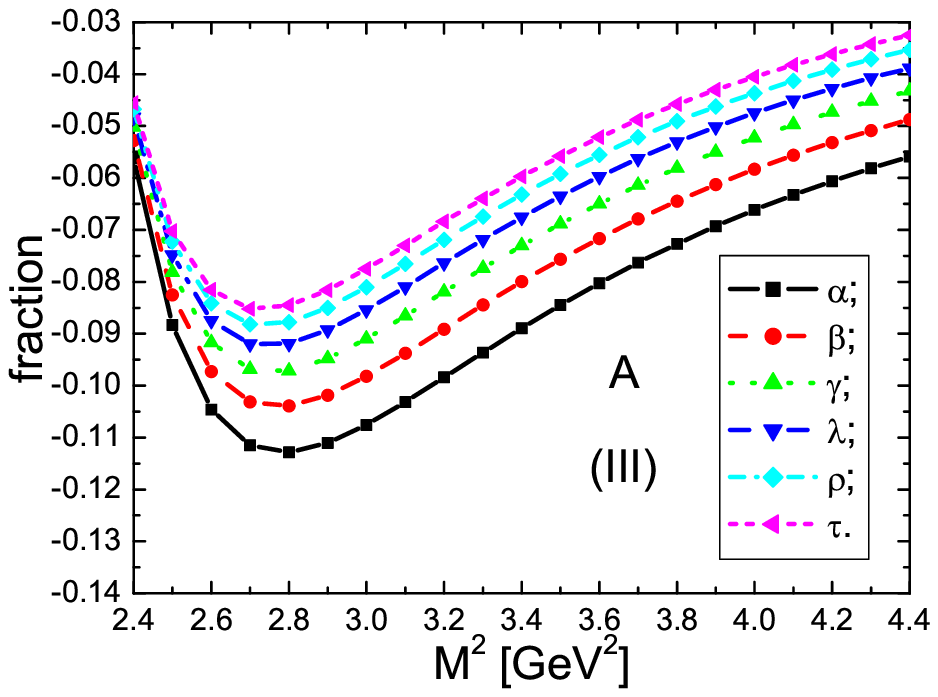}
 \includegraphics[totalheight=5cm,width=6cm]{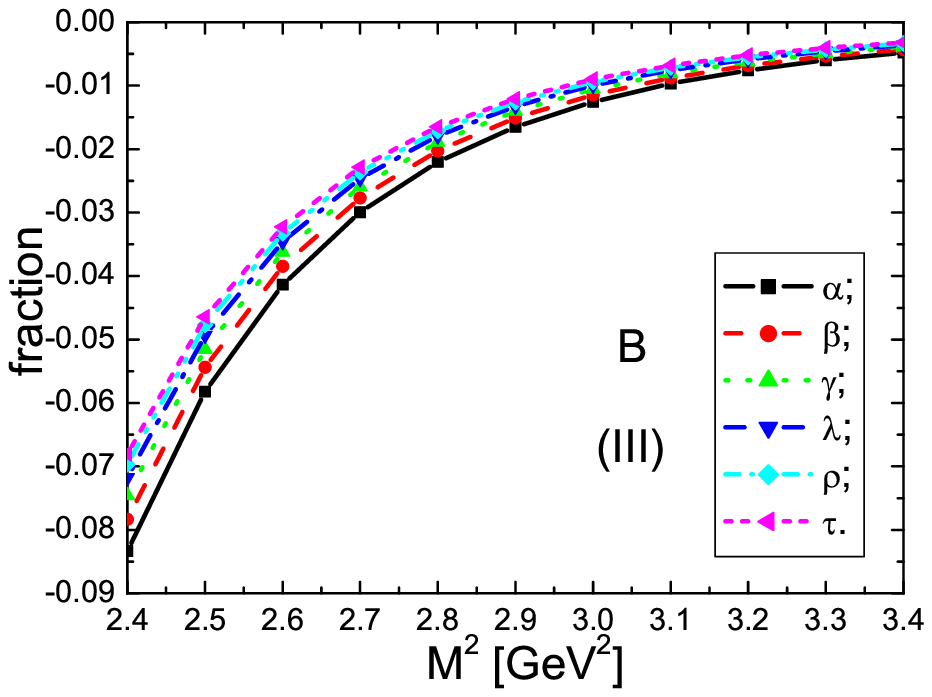}
   \caption{ The contributions from different terms with variation of the Borel
   parameter $M^2$  in the operator product expansion for the $C\gamma_5-C\gamma_\mu \gamma_5$ type current operators. The $A$ and
   $B$ denote the contributions from the
   $\langle \bar{q}q\rangle^2+\langle \bar{q}q\rangle \langle \bar{q}g_s \sigma Gq\rangle
   $ term and the  $ \langle \bar{q}g_s \sigma Gq\rangle^2
   $ term,  respectively. The
   (I), (II) and (III) denote the $c\bar{c}q\bar{q}$,
   $c\bar{c}q\bar{s}$ and $c\bar{c}s\bar{s}$ channels, respectively. The notations
   $\alpha$, $\beta$, $\gamma$, $\lambda$, $\rho$ and $\tau$  correspond to the threshold
   parameters $s_0=28\,\rm{GeV}^2$,
   $29\,\rm{GeV}^2$, $30\,\rm{GeV}^2$, $31\,\rm{GeV}^2$, $32\,\rm{GeV}^2$ and $33\,\rm{GeV}^2$, respectively.}
\end{figure}

\begin{figure}
 \centering
 \includegraphics[totalheight=5cm,width=6cm]{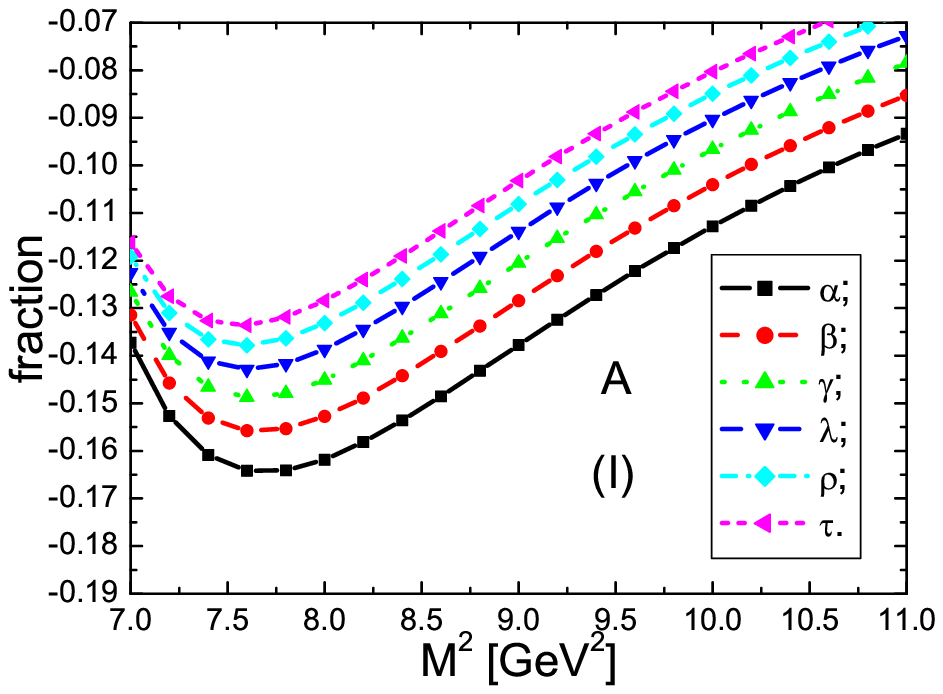}
 \includegraphics[totalheight=5cm,width=6cm]{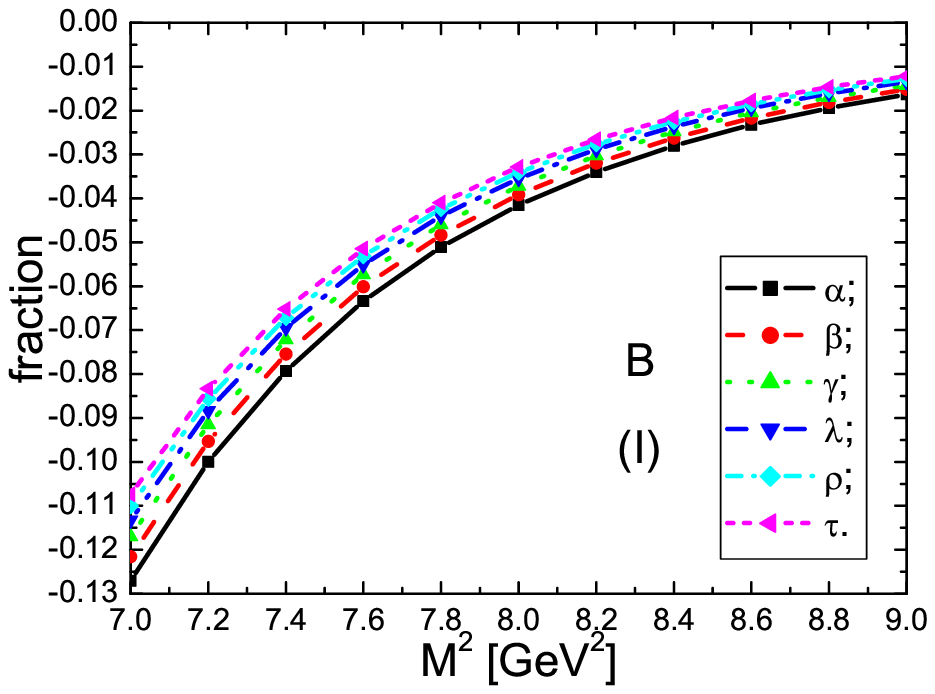}
 \includegraphics[totalheight=5cm,width=6cm]{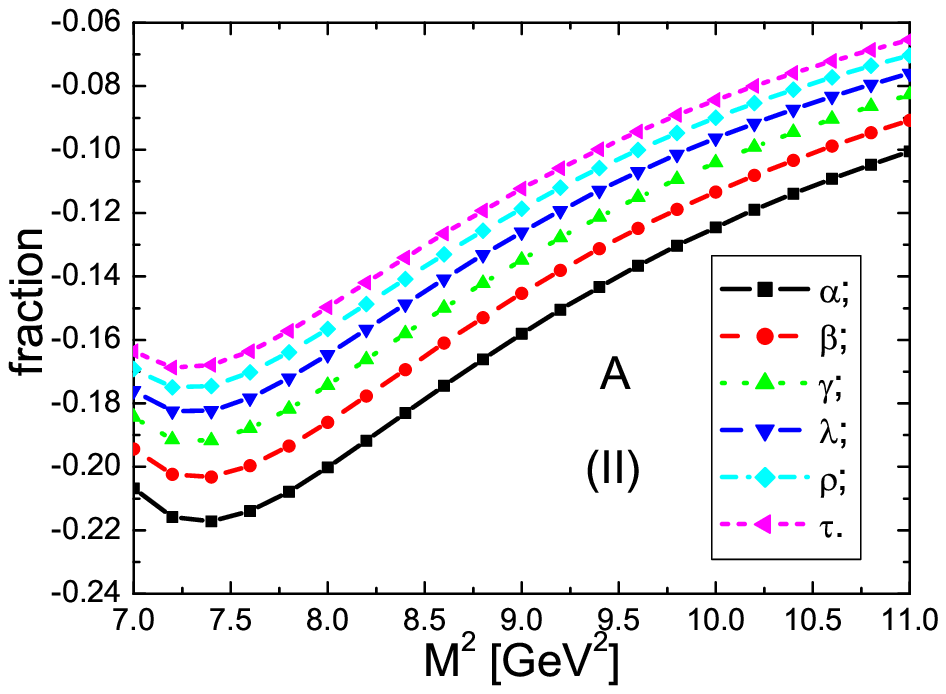}
 \includegraphics[totalheight=5cm,width=6cm]{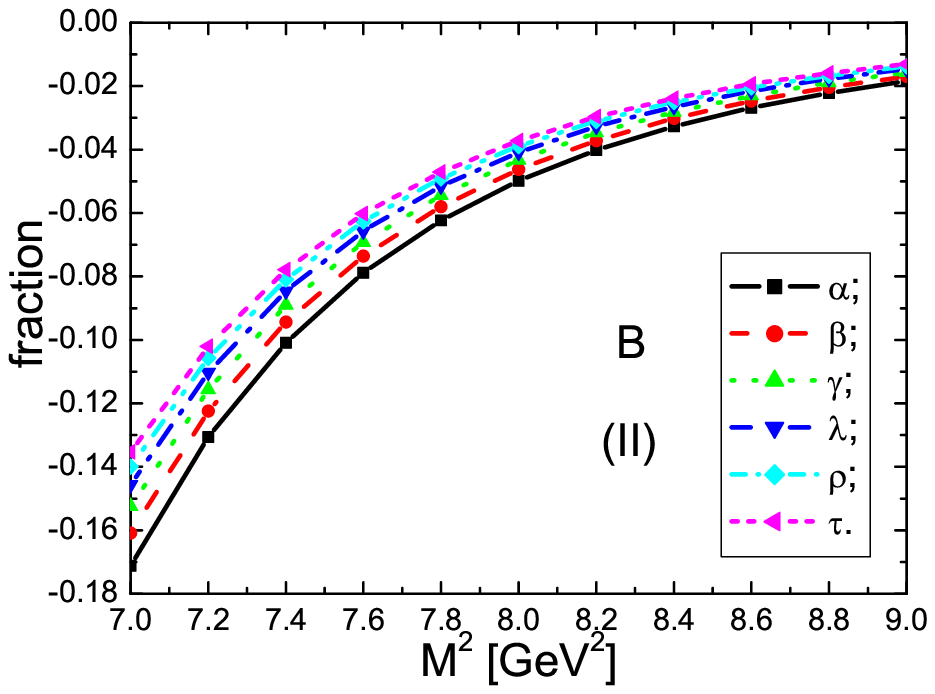}
 \includegraphics[totalheight=5cm,width=6cm]{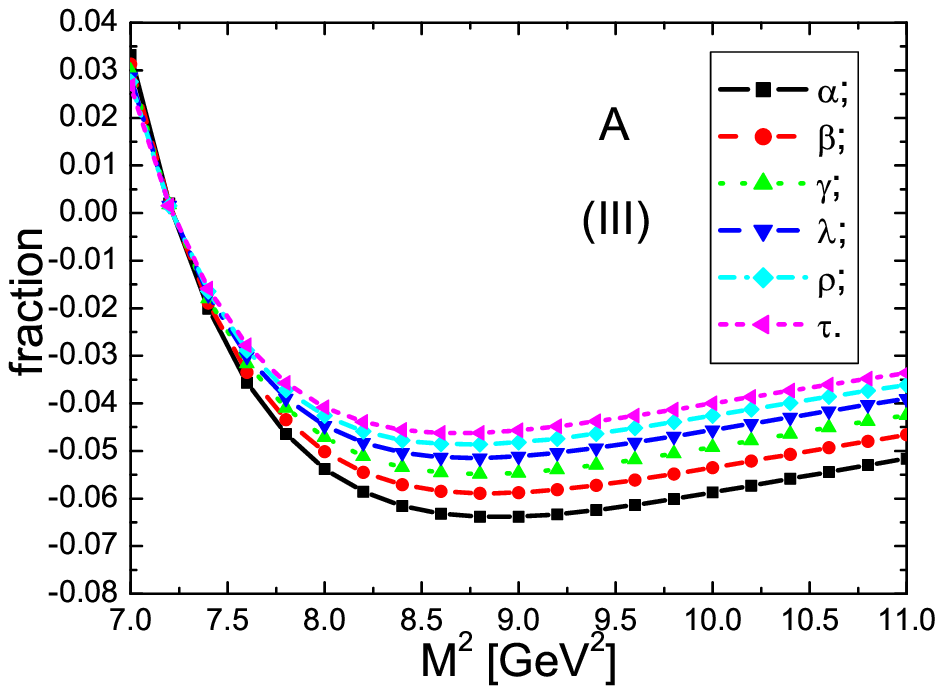}
 \includegraphics[totalheight=5cm,width=6cm]{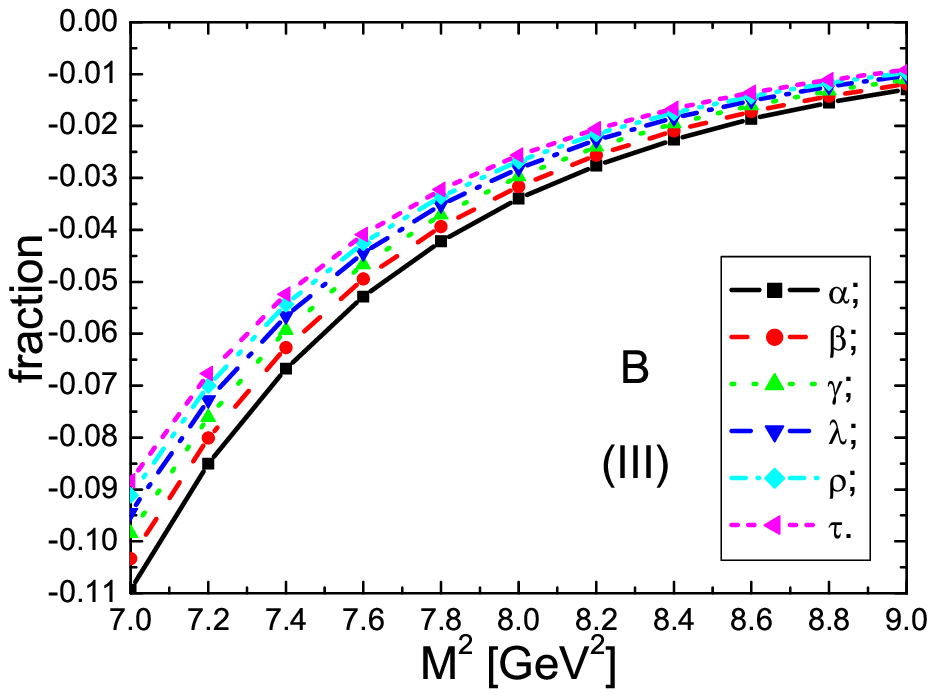}
   \caption{ The contributions from different terms with variation of the Borel
   parameter $M^2$  in the operator product expansion for the $C\gamma_5-C\gamma_\mu \gamma_5$ type current operators.
The $A$ and
   $B$ denote the contributions from the
   $\langle \bar{q}q\rangle^2+\langle \bar{q}q\rangle \langle \bar{q}g_s \sigma Gq\rangle
   $ term and the  $ \langle \bar{q}g_s \sigma Gq\rangle^2
   $ term,  respectively.   The
   (I), (II) and (III) denote the $b\bar{b}q\bar{q}$,
   $b\bar{b}q\bar{s}$ and $b\bar{b}s\bar{s}$ channels, respectively. The notations
   $\alpha$, $\beta$, $\gamma$, $\lambda$, $\rho$ and $\tau$  correspond to the threshold
   parameters $s_0=152\,\rm{GeV}^2$,
   $154\,\rm{GeV}^2$, $156\,\rm{GeV}^2$, $158\,\rm{GeV}^2$, $160\,\rm{GeV}^2$ and $162\,\rm{GeV}^2$, respectively. }
\end{figure}

\begin{figure}
 \centering
 \includegraphics[totalheight=5cm,width=6cm]{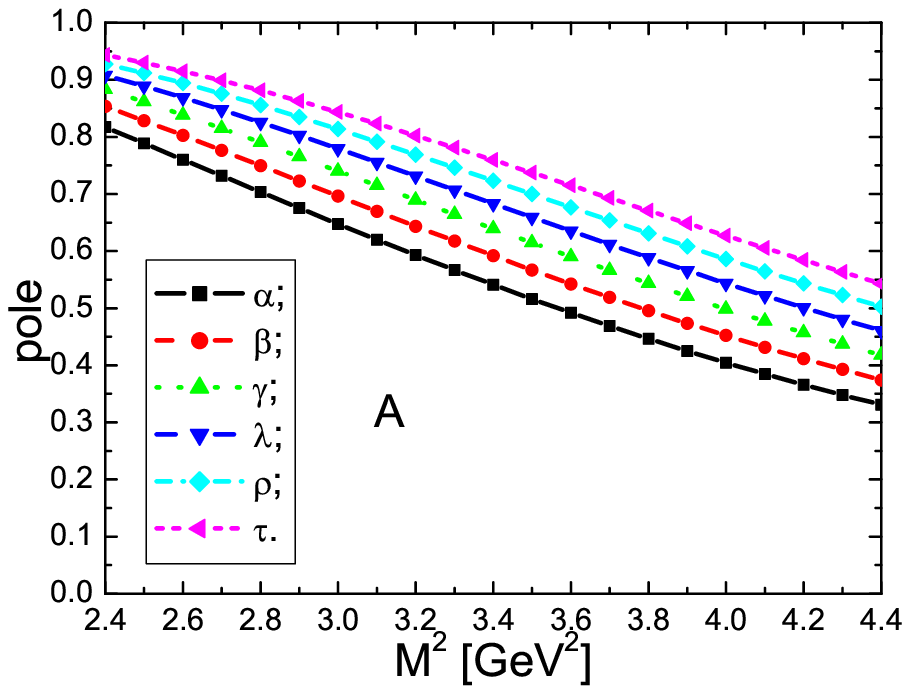}
 \includegraphics[totalheight=5cm,width=6cm]{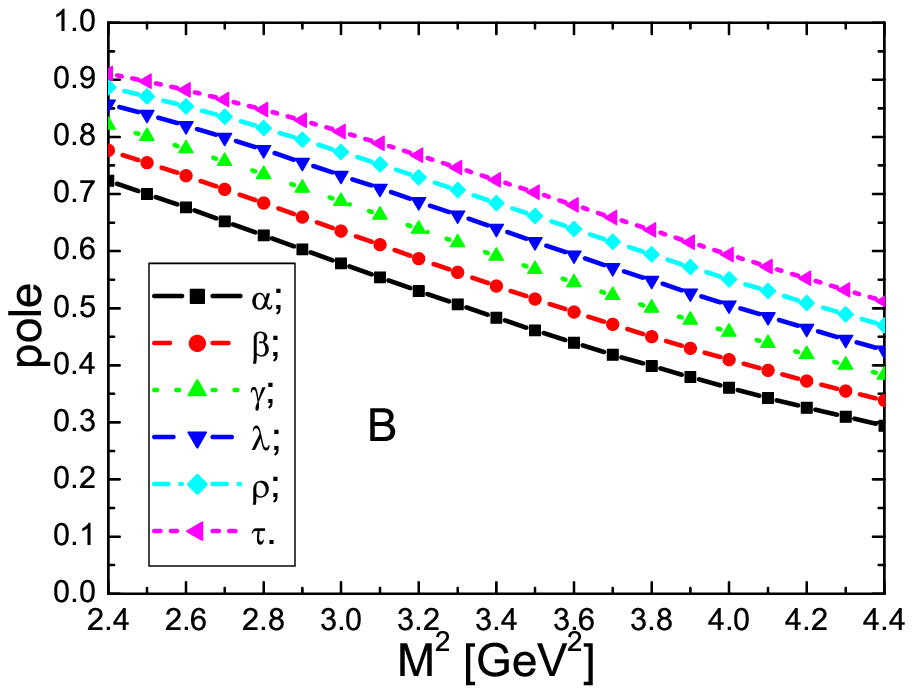}
 \includegraphics[totalheight=5cm,width=6cm]{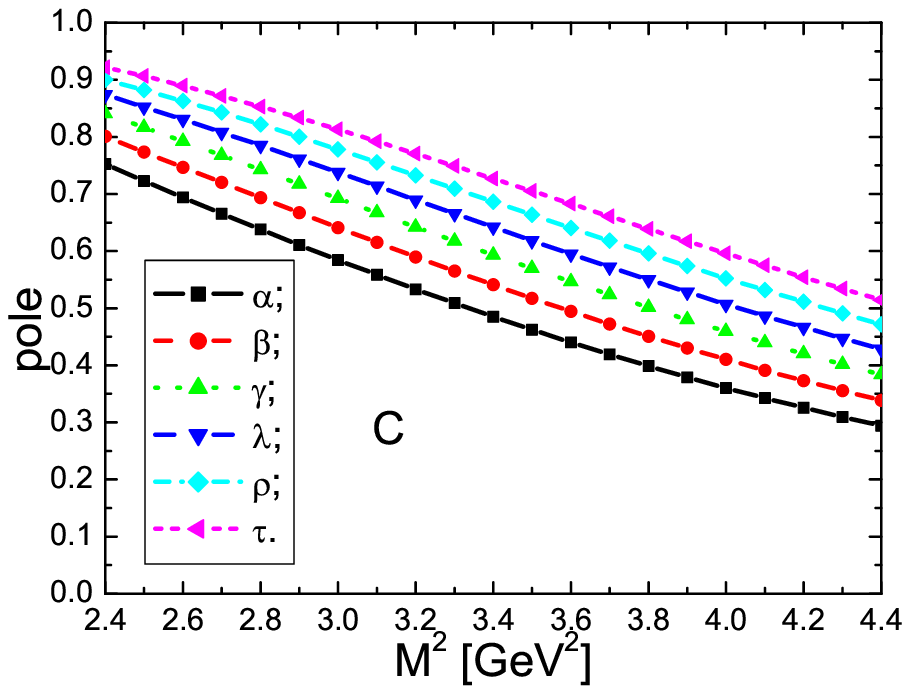}
 \includegraphics[totalheight=5cm,width=6cm]{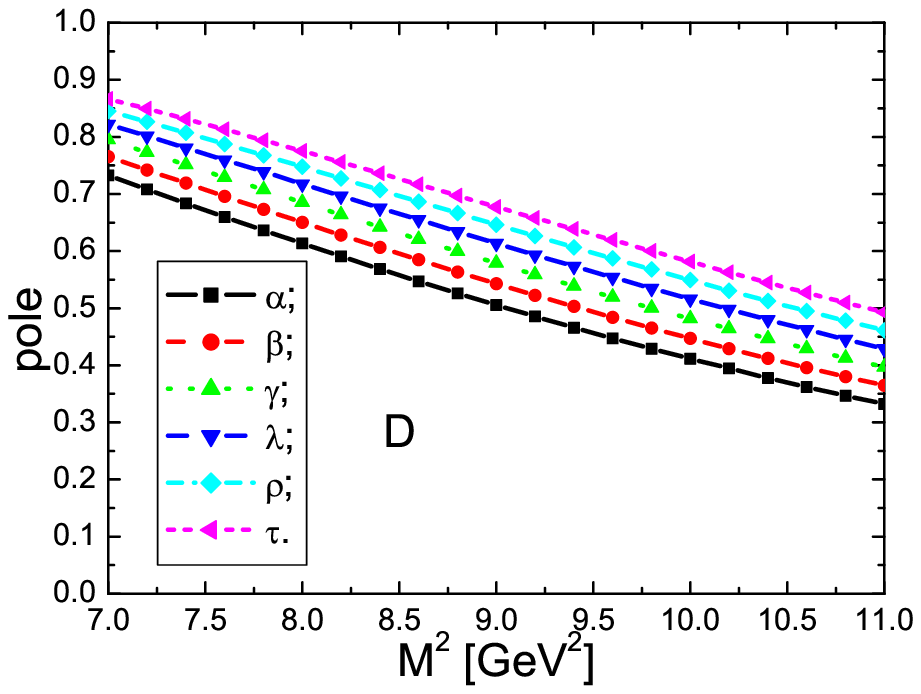}
 \includegraphics[totalheight=5cm,width=6cm]{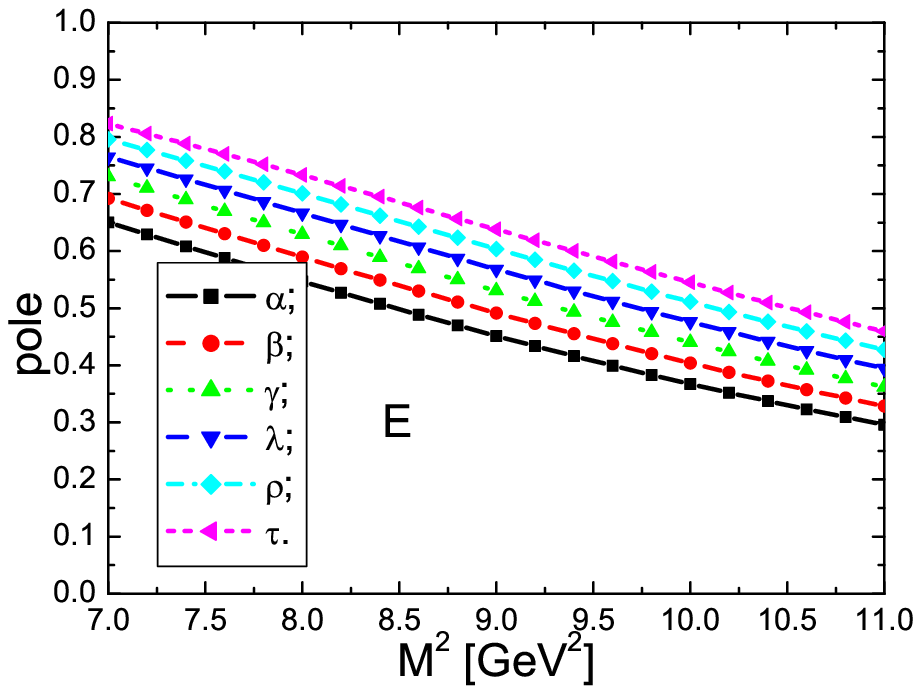}
 \includegraphics[totalheight=5cm,width=6cm]{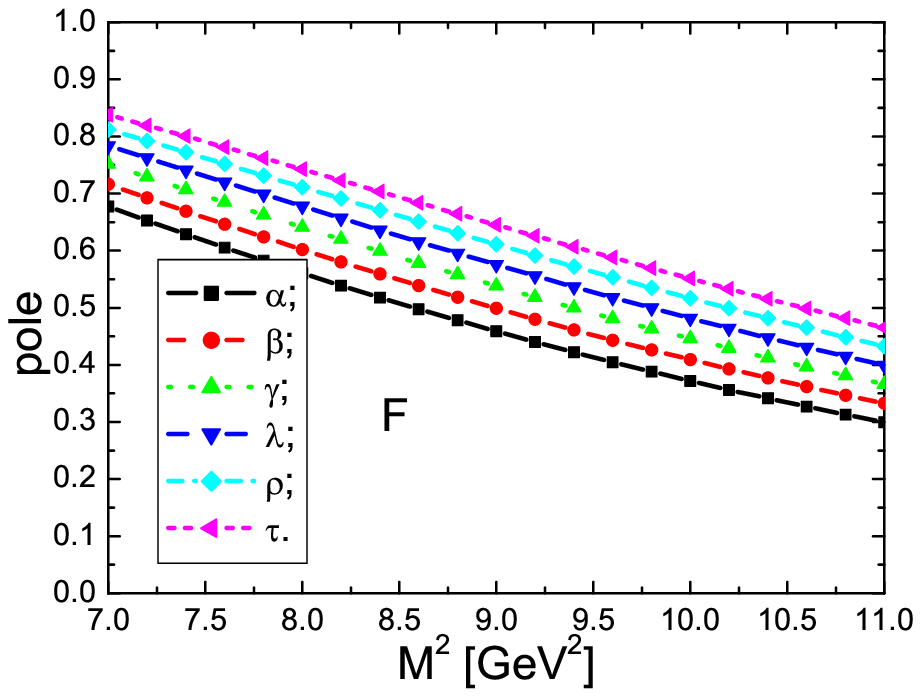}
   \caption{ The contributions from the pole terms with variation of the Borel parameter $M^2$ for the $C\gamma_5-C\gamma_\mu \gamma_5$ type current opertors. The $A$, $B$, $C$,
   $D$, $E$ and $F$ denote the $c\bar{c}q\bar{q}$,
   $c\bar{c}q\bar{s}$, $c\bar{c}s\bar{s}$, $b\bar{b}q\bar{q}$,
   $b\bar{b}q\bar{s}$ and $b\bar{b}s\bar{s}$ channels, respectively.   In the $c\bar{c}$ channels, the notations
   $\alpha$, $\beta$, $\gamma$, $\lambda$, $\rho$ and $\tau$  correspond to the threshold
   parameters $s_0=28\,\rm{GeV}^2$,
   $29\,\rm{GeV}^2$, $30\,\rm{GeV}^2$, $31\,\rm{GeV}^2$, $32\,\rm{GeV}^2$ and $33\,\rm{GeV}^2$ respectively
   ;  while in the $b\bar{b}$ channels they correspond to
    the threshold
   parameters  $s_0=152\,\rm{GeV}^2$,
   $154\,\rm{GeV}^2$, $156\,\rm{GeV}^2$, $158\,\rm{GeV}^2$, $160\,\rm{GeV}^2$ and $162\,\rm{GeV}^2$ respectively. }
\end{figure}

Taking into account all uncertainties of the input parameters,
finally we obtain the values of the masses and pole resides of
 the vector hidden charm and bottom tetraquark states  $Z$, which are  shown in Figs.4-7 and Tables 1-2.

\begin{figure}
 \centering
 \includegraphics[totalheight=5cm,width=6cm]{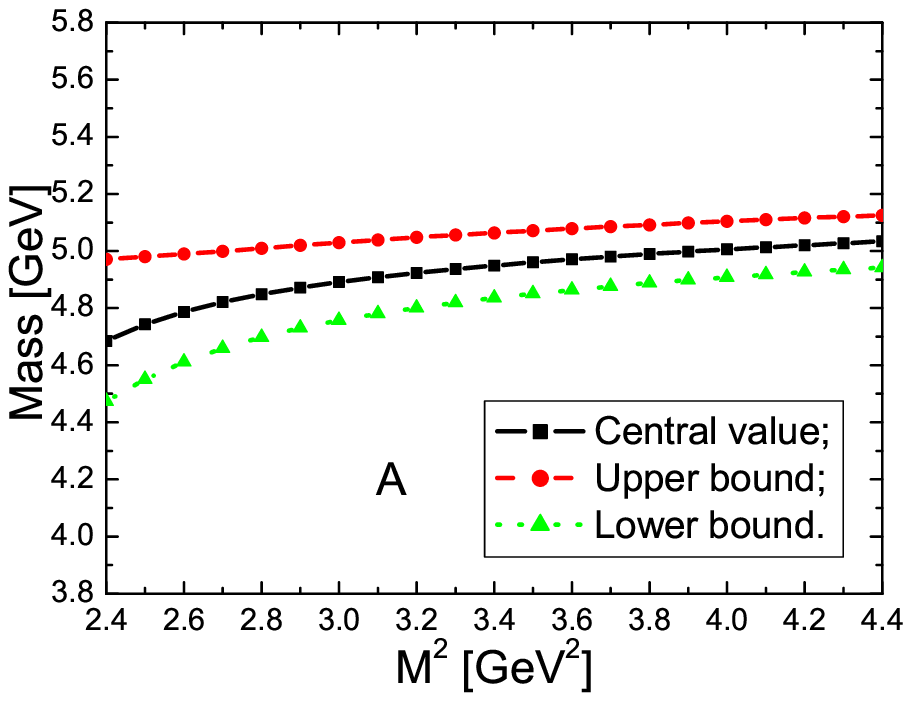}
 \includegraphics[totalheight=5cm,width=6cm]{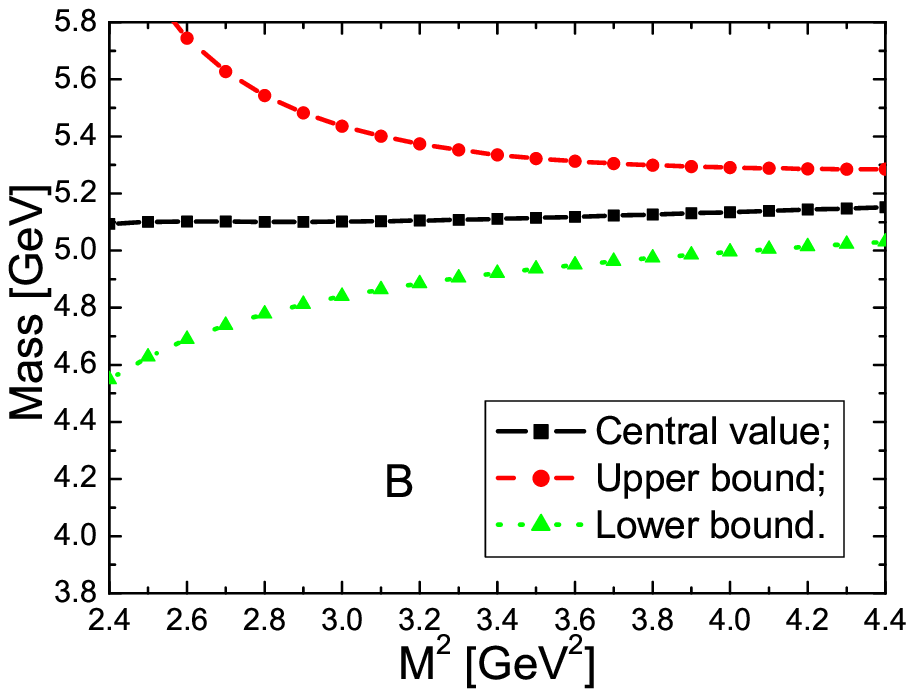}
 \includegraphics[totalheight=5cm,width=6cm]{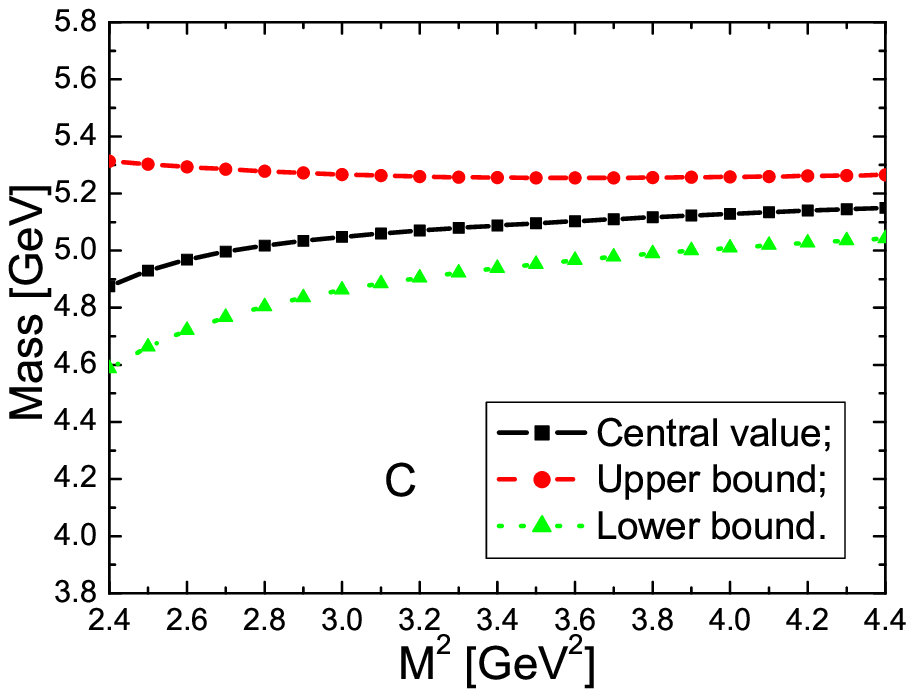}
 \includegraphics[totalheight=5cm,width=6cm]{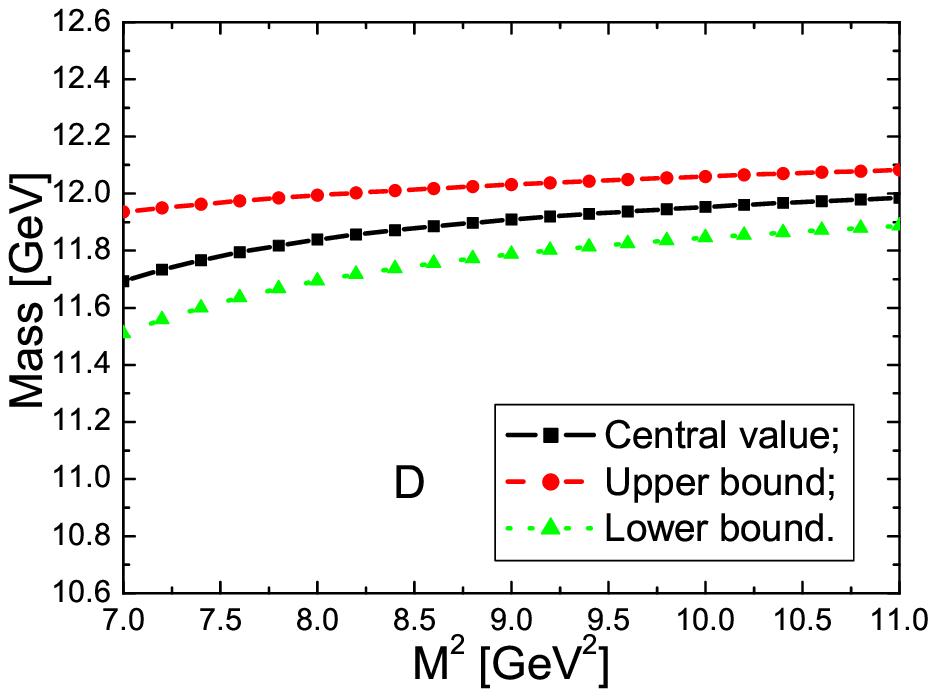}
 \includegraphics[totalheight=5cm,width=6cm]{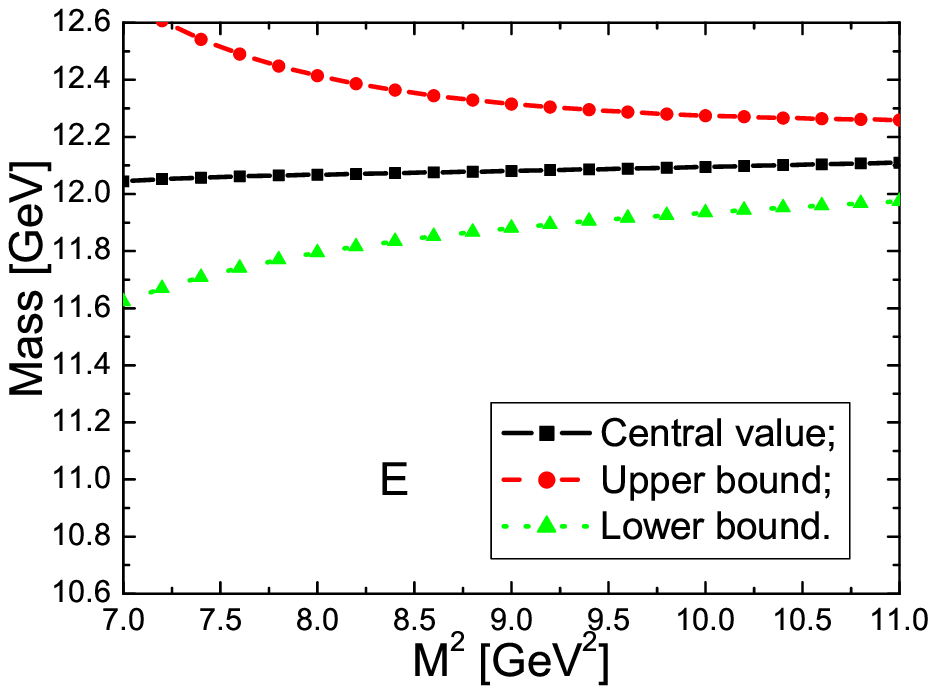}
 \includegraphics[totalheight=5cm,width=6cm]{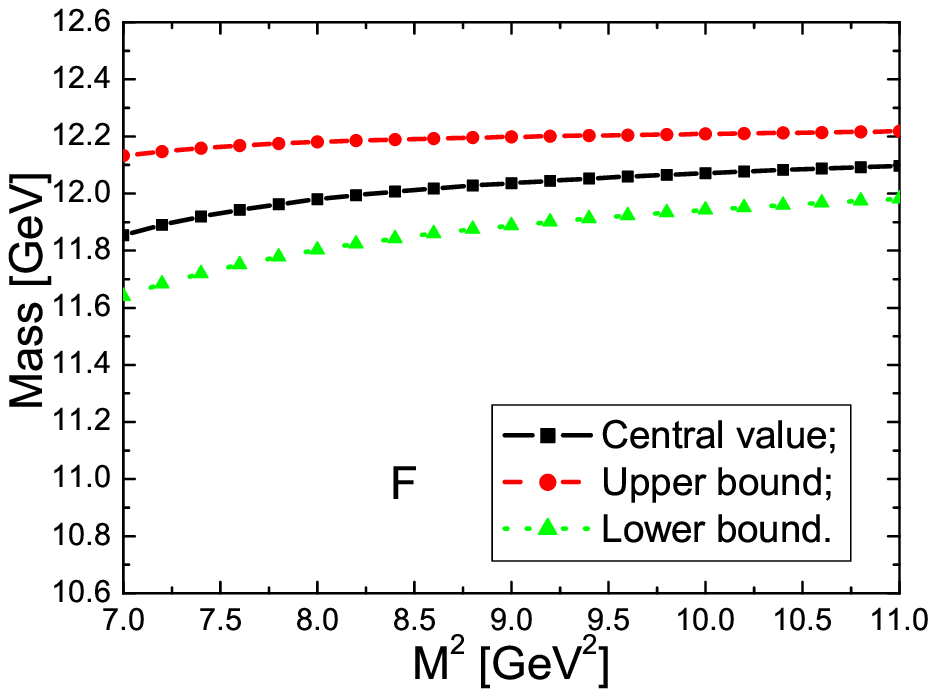}
   \caption{ The masses of the vector tetraquark states with variation of the Borel parameter $M^2$ for the $C\gamma_5-C\gamma_\mu \gamma_5$ type current opertors. The $A$, $B$, $C$,
   $D$, $E$ and $F$ denote the $c\bar{c}q\bar{q}$,
   $c\bar{c}q\bar{s}$, $c\bar{c}s\bar{s}$, $b\bar{b}q\bar{q}$,
   $b\bar{b}q\bar{s}$ and $b\bar{b}s\bar{s}$ channels, respectively. }
\end{figure}

\begin{figure}
 \centering
 \includegraphics[totalheight=5cm,width=6cm]{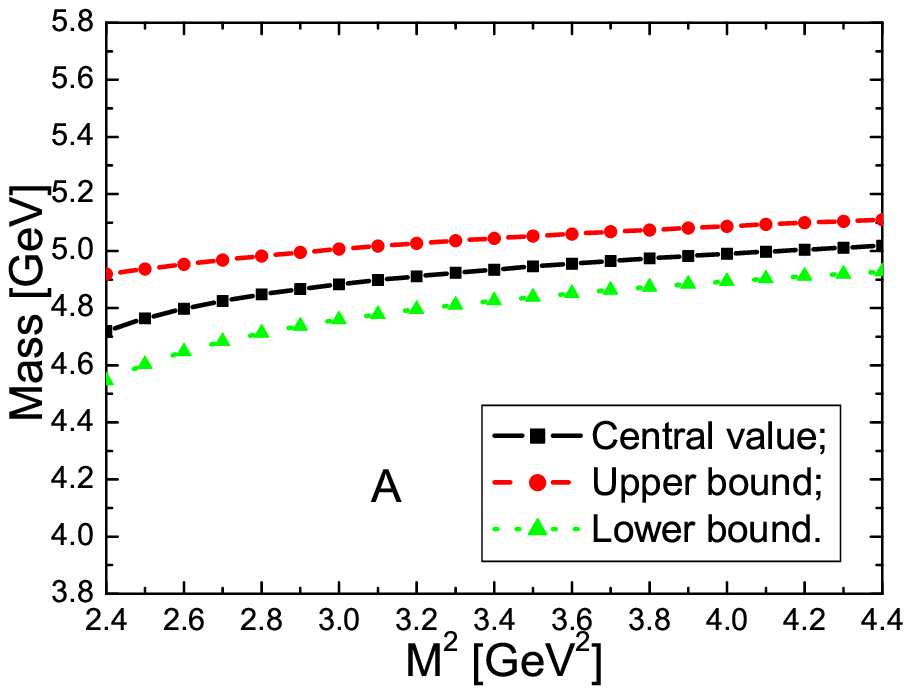}
 \includegraphics[totalheight=5cm,width=6cm]{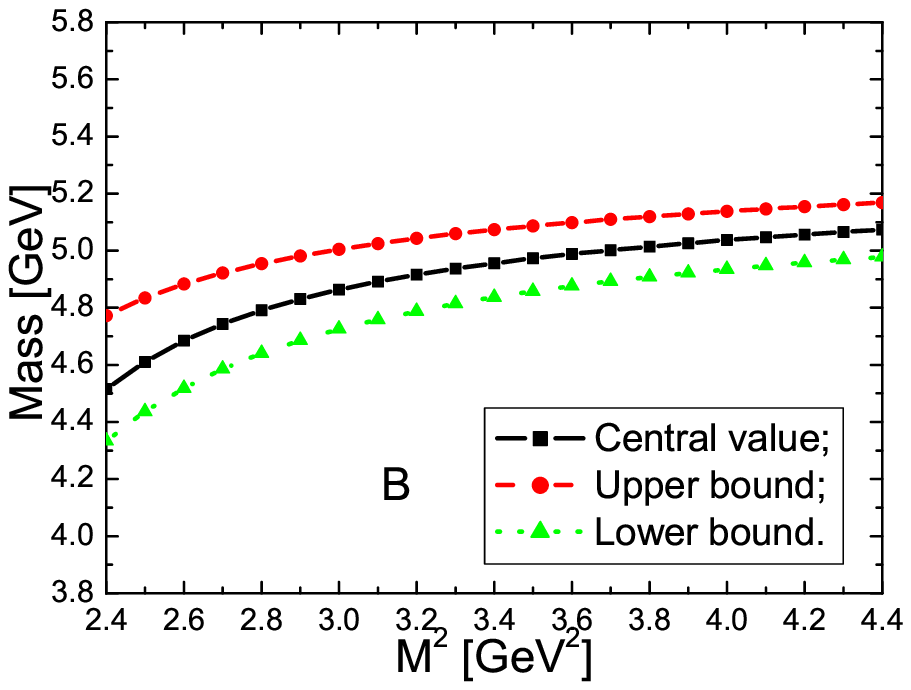}
 \includegraphics[totalheight=5cm,width=6cm]{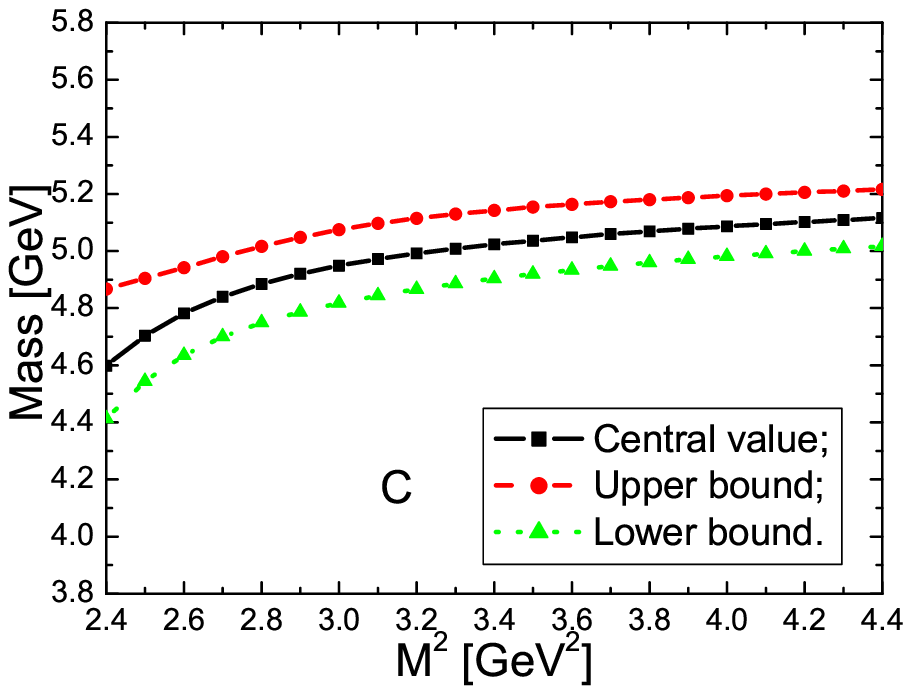}
 \includegraphics[totalheight=5cm,width=6cm]{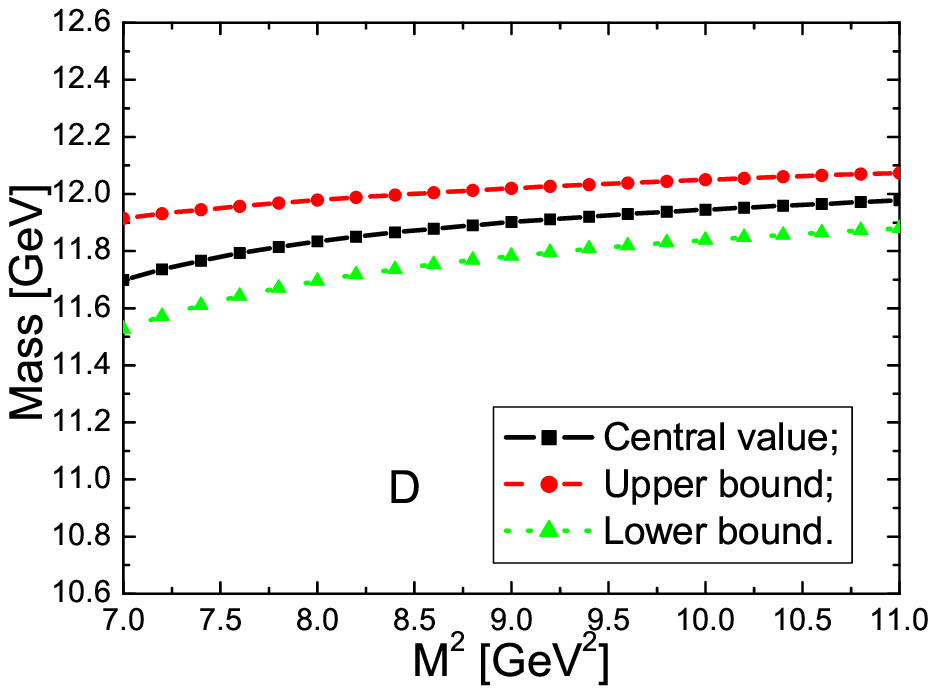}
 \includegraphics[totalheight=5cm,width=6cm]{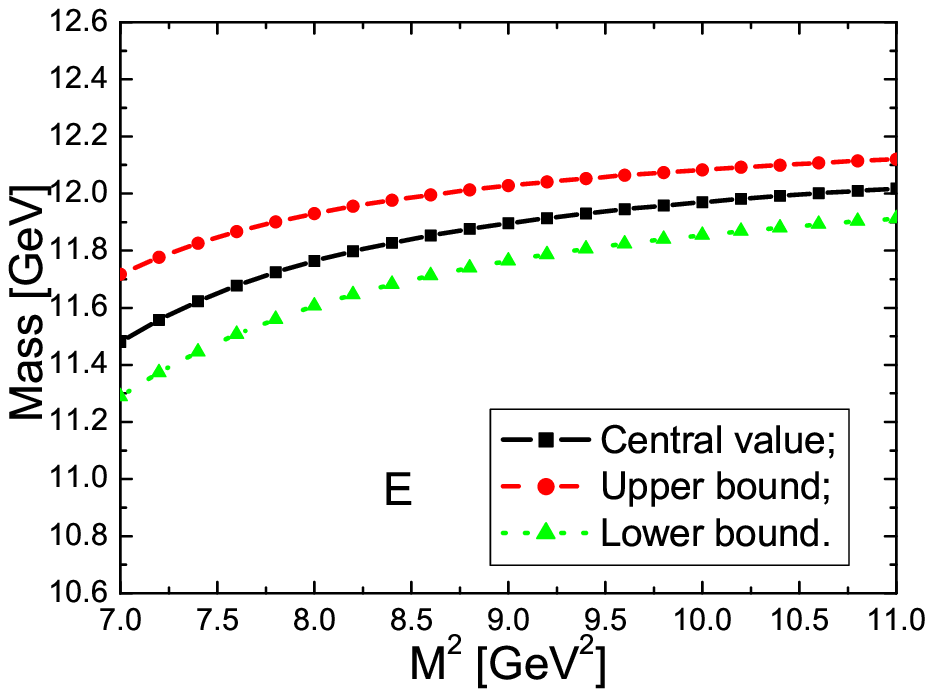}
 \includegraphics[totalheight=5cm,width=6cm]{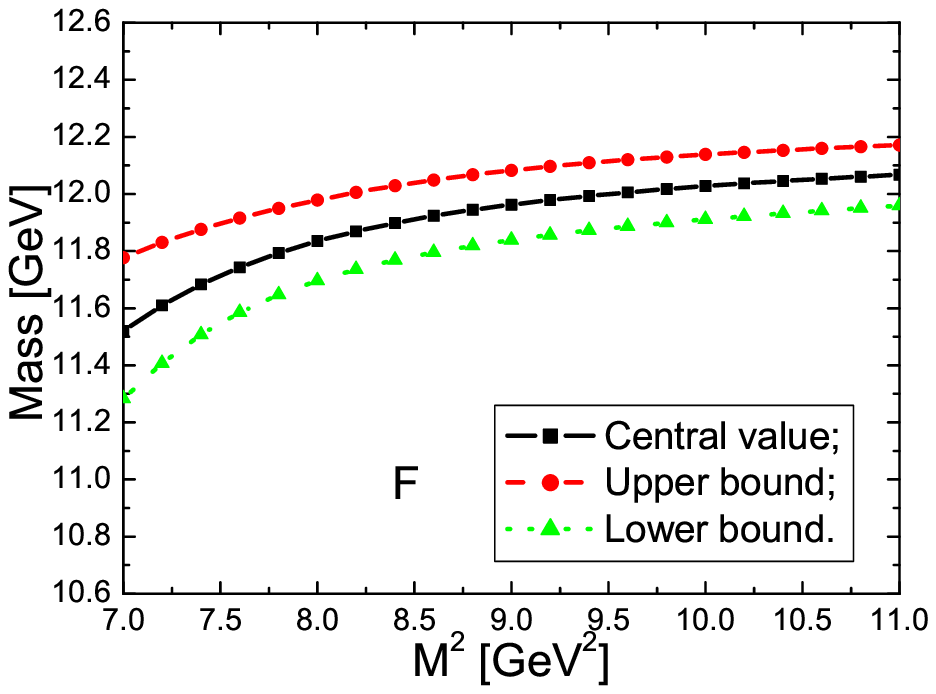}
   \caption{ The masses of the vector tetraquark states with variation of the Borel parameter $M^2$ for the
   $C-C\gamma_\mu $ type current opertors. The $A$, $B$, $C$,
   $D$, $E$ and $F$ denote the $c\bar{c}q\bar{q}$,
   $c\bar{c}q\bar{s}$, $c\bar{c}s\bar{s}$, $b\bar{b}q\bar{q}$,
   $b\bar{b}q\bar{s}$ and $b\bar{b}s\bar{s}$ channels, respectively. }
\end{figure}

\begin{figure}
 \centering
 \includegraphics[totalheight=5cm,width=6cm]{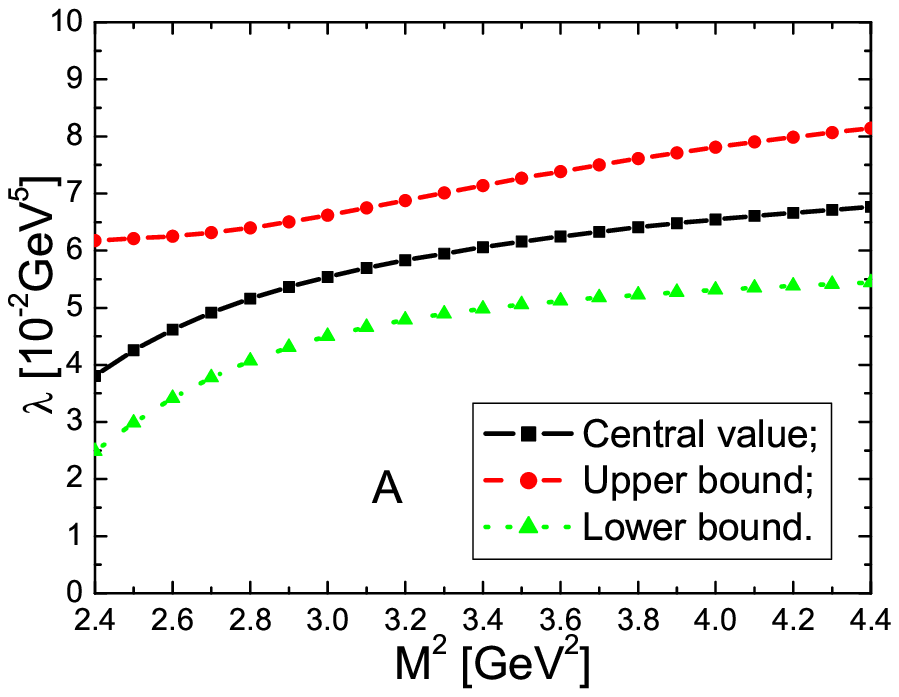}
 \includegraphics[totalheight=5cm,width=6cm]{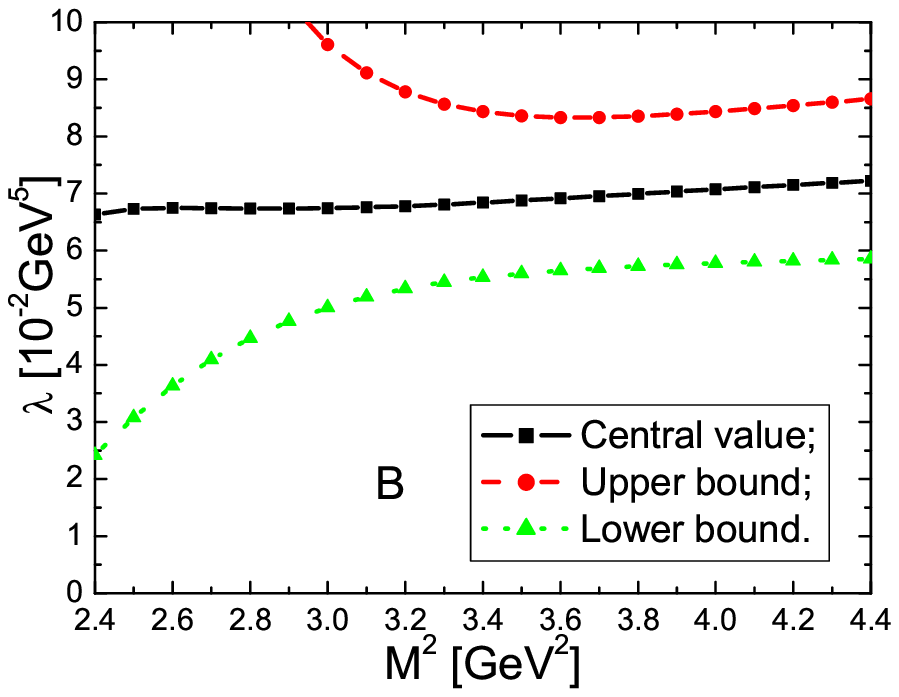}
 \includegraphics[totalheight=5cm,width=6cm]{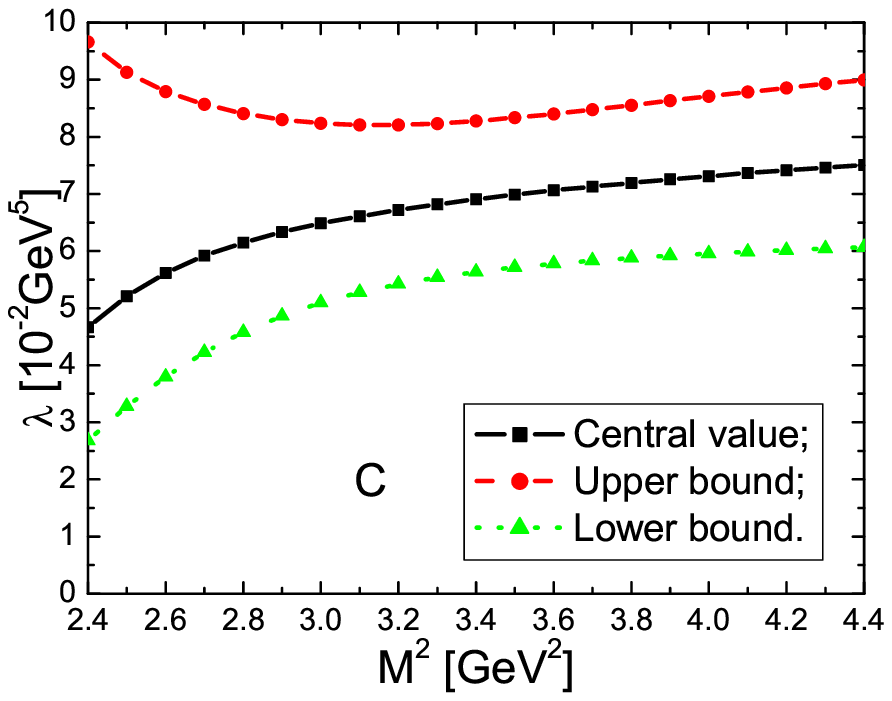}
 \includegraphics[totalheight=5cm,width=6cm]{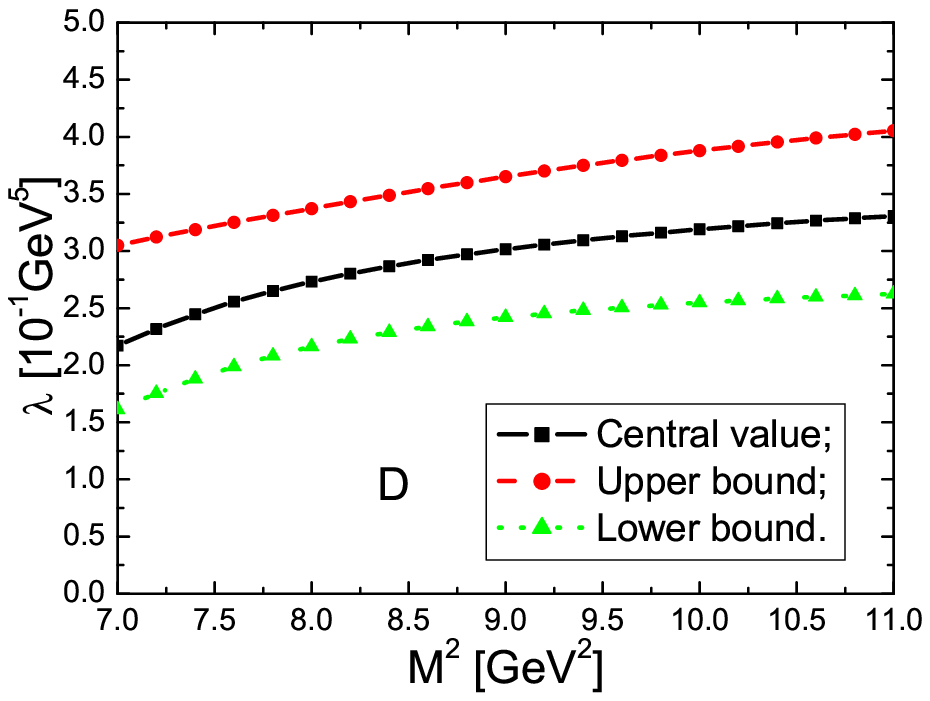}
 \includegraphics[totalheight=5cm,width=6cm]{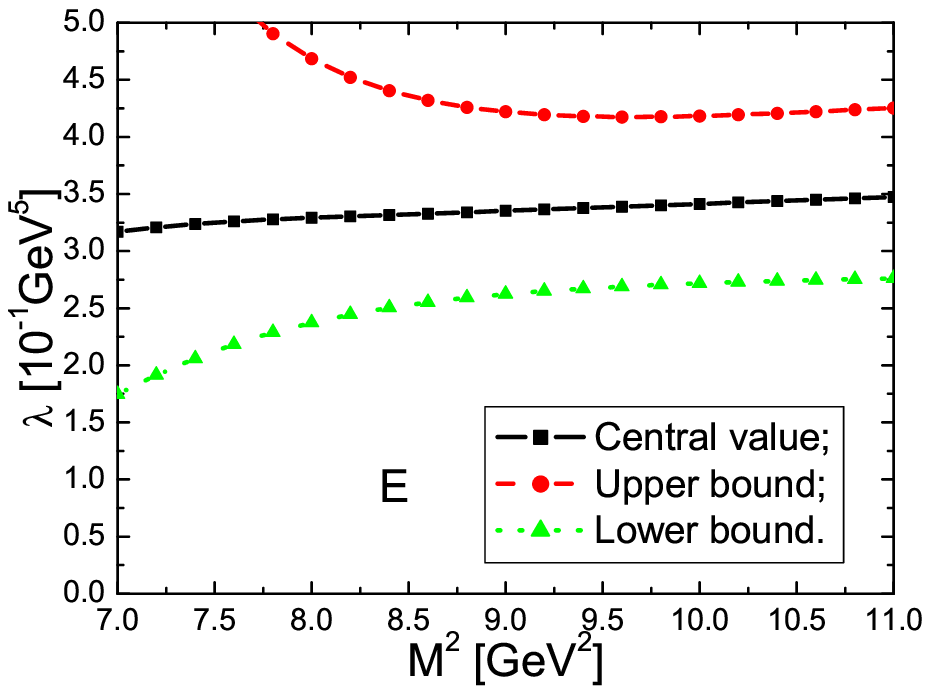}
 \includegraphics[totalheight=5cm,width=6cm]{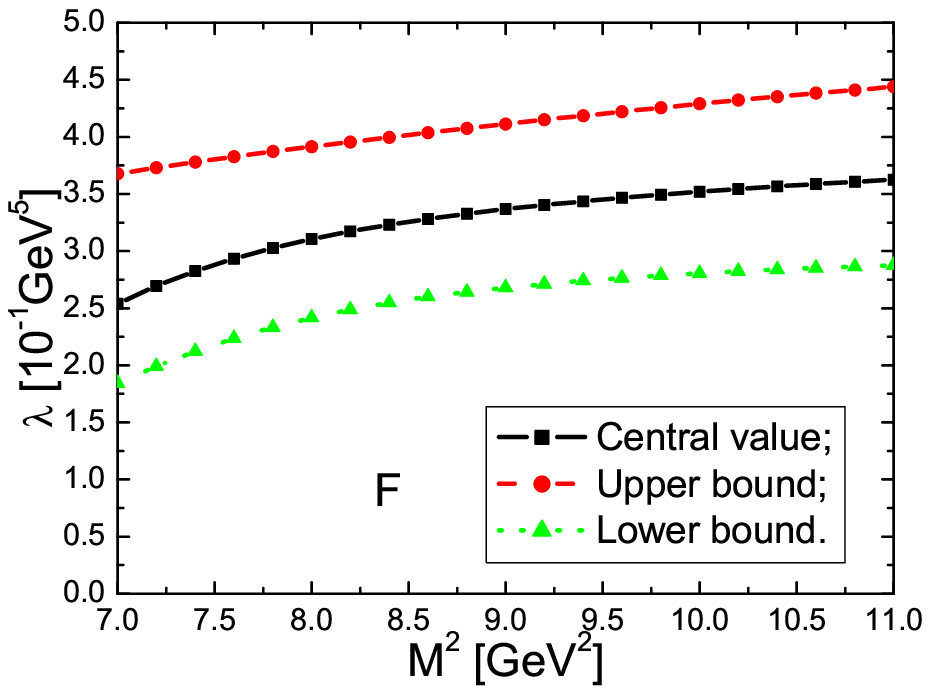}
   \caption{ The pole residues of the vector tetraquark states with variation of the Borel parameter $M^2$ for the $C\gamma_5-C\gamma_\mu \gamma_5$ type current opertors. The $A$, $B$, $C$,
   $D$, $E$ and $F$ denote the $c\bar{c}q\bar{q}$,
   $c\bar{c}q\bar{s}$, $c\bar{c}s\bar{s}$, $b\bar{b}q\bar{q}$,
   $b\bar{b}q\bar{s}$ and $b\bar{b}s\bar{s}$ channels, respectively.}
\end{figure}

\begin{figure}
 \centering
 \includegraphics[totalheight=5cm,width=6cm]{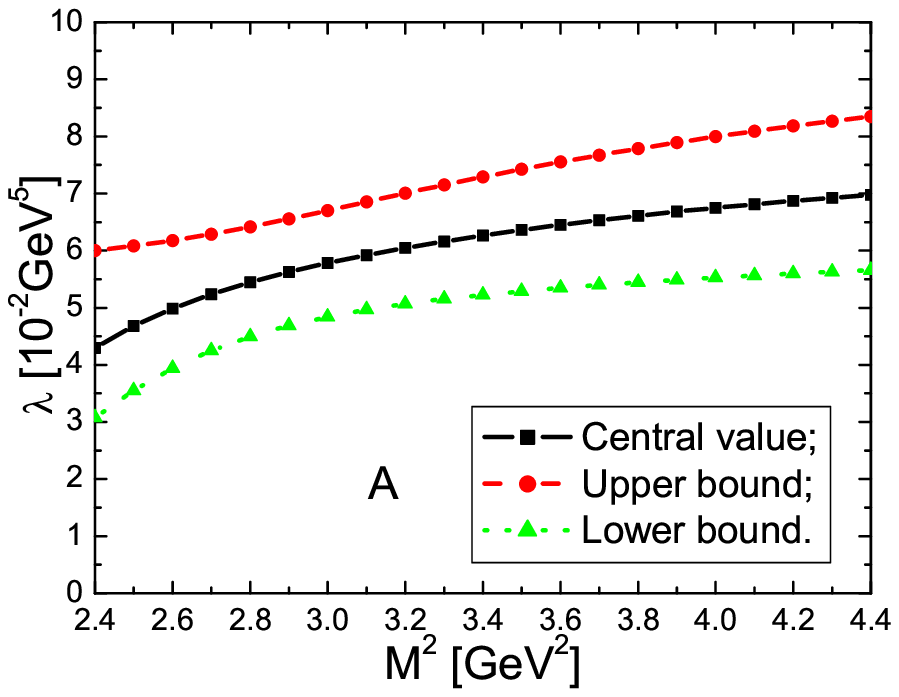}
 \includegraphics[totalheight=5cm,width=6cm]{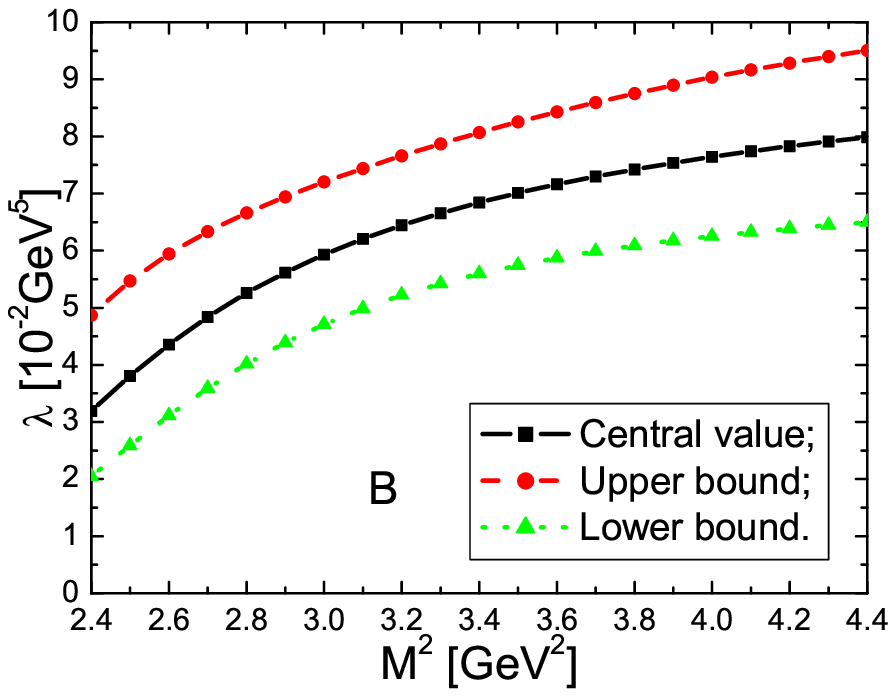}
 \includegraphics[totalheight=5cm,width=6cm]{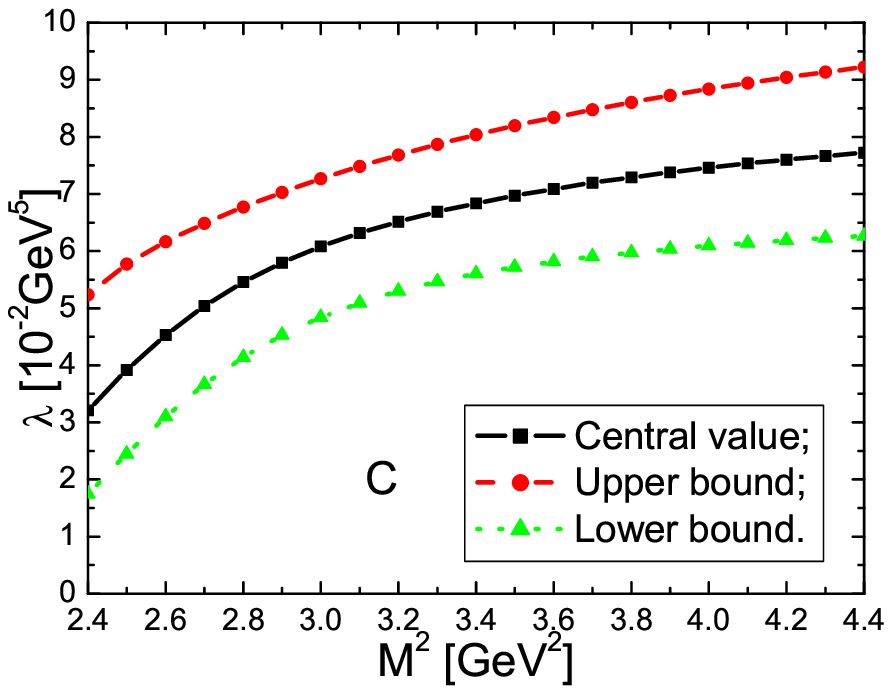}
 \includegraphics[totalheight=5cm,width=6cm]{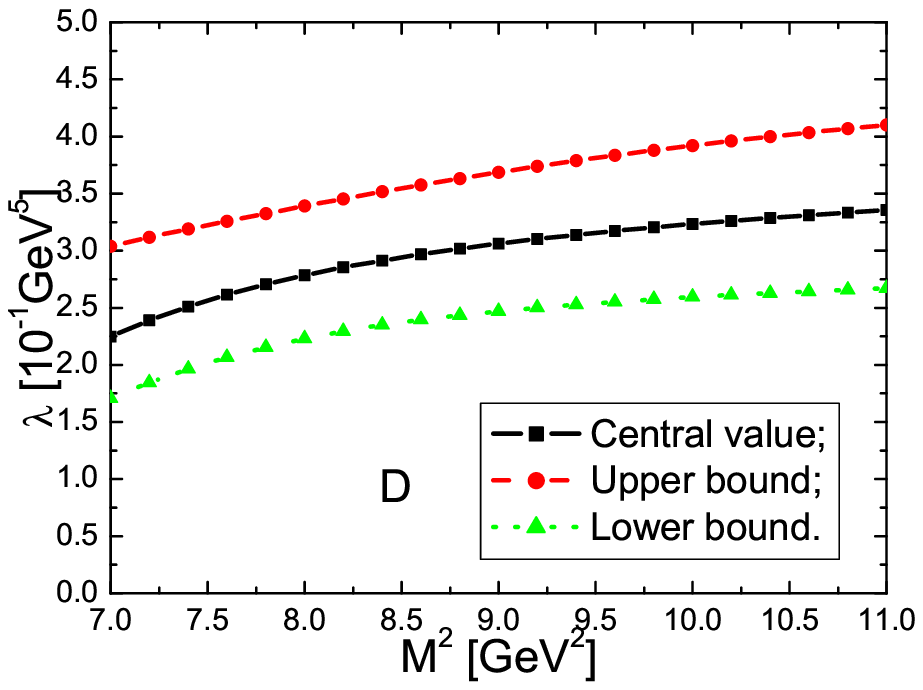}
 \includegraphics[totalheight=5cm,width=6cm]{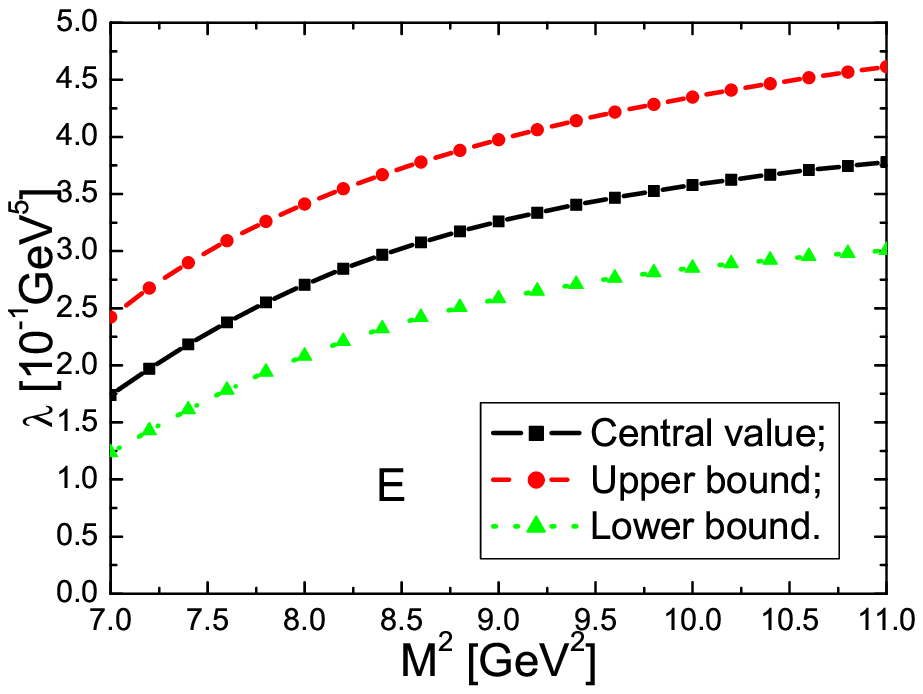}
 \includegraphics[totalheight=5cm,width=6cm]{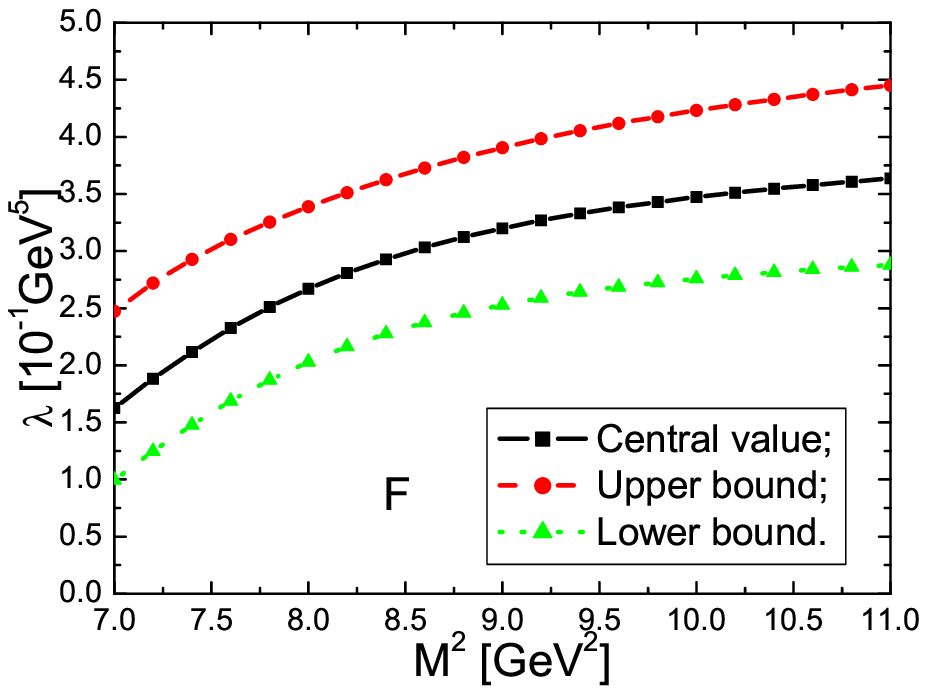}
   \caption{ The pole residues of the vector tetraquark states with variation of the Borel parameter $M^2$ for the
   $C-C\gamma_\mu $ type current opertors. The $A$, $B$, $C$,
   $D$, $E$ and $F$ denote the $c\bar{c}q\bar{q}$,
   $c\bar{c}q\bar{s}$, $c\bar{c}s\bar{s}$, $b\bar{b}q\bar{q}$,
   $b\bar{b}q\bar{s}$ and $b\bar{b}s\bar{s}$ channels, respectively. }
\end{figure}

\begin{table}
\begin{center}
\begin{tabular}{|c|c|c|c|}
\hline\hline tetraquark states & $C\gamma_5-C\gamma_\mu\gamma_5$ & $C-C\gamma_\mu$\\
\hline
      $c\bar{c}q\bar{q}$  &$4.97\pm0.14$ &$4.95\pm0.14$\\ \hline
       $ c\bar{c}q\bar{s}$& $5.12\pm0.20$& $5.00\pm0.15$\\     \hline
      $c\bar{c}s\bar{s} $ &$5.10\pm0.17$& $5.05\pm0.15$\\      \hline
    $b\bar{b}q\bar{q}$  &$11.90\pm0.14$ &$11.90\pm0.14$\\ \hline
       $ b\bar{b}q\bar{s}$& $12.08\pm0.23$ &$11.89\pm0.17$\\     \hline
      $ b\bar{b}s\bar{s} $ &$12.04\pm0.17$ &$11.95\pm0.16$\\      \hline
    \hline
\end{tabular}
\end{center}
\caption{ The masses (in unit of GeV)  of the vector tetraquark
states. }
\end{table}

\begin{table}
\begin{center}
\begin{tabular}{|c|c|c|c|}
\hline\hline tetraquark states & $C\gamma_5-C\gamma_\mu\gamma_5$ & $C-C\gamma_\mu$\\
\hline
      $c\bar{c}q\bar{q}$  &$6.3\pm1.4$ &$6.5\pm1.4$\\ \hline
       $ c\bar{c}q\bar{s}$& $6.9\pm1.4$& $7.2\pm1.7$\\     \hline
      $c\bar{c}s\bar{s} $ &$7.1\pm1.5$& $7.1\pm1.6$\\      \hline
    $b\bar{b}q\bar{q}$  &$3.0\pm0.7$ &$3.1\pm0.7$\\ \hline
       $ b\bar{b}q\bar{s}$& $3.4\pm0.8$ &$3.3\pm0.8$\\     \hline
      $ b\bar{b}s\bar{s} $ &$3.4\pm0.8$ &$3.2\pm0.8$\\      \hline
    \hline
\end{tabular}
\end{center}
\caption{ The  pole residues (in unit of $10^{-2}\, \rm{GeV}^5$ and
$10^{-1}\, \rm{GeV}^5$ for the $c\bar{c}$ and $b\bar{b}$ channels
respectively) of the vector tetraquark states. }
\end{table}

From Table 1, we can see that the $SU(3)$ breaking effects for the
masses of the hidden charm and bottom tetraquark states are buried
in the uncertainties. The central values of the vector tetraquark
state $c\bar{c}q\bar{q}$ is slightly below the ones
 $M_{Z}=(5.12\pm0.15)\,\rm{GeV}$ and
$M_{Z}=(5.16\pm0.16)\,\rm{GeV}$  obtained in Ref.\cite{Wang0807},
about $0.15\,\rm{GeV}$. In  Ref.\cite{Wang0807}, the contributions
from the terms $\langle \bar{q}q\rangle \langle \bar{q}g_s \sigma
Gq\rangle
   $  and   $ \langle \bar{q}g_s \sigma Gq\rangle^2
   $  are  neglected.

We calculate the mass spectrum of the vector hidden charm and bottom
tetraquark states  by imposing the two criteria of the QCD sum
rules. In fact, we usually  consult the experimental data in
choosing the Borel parameter $M^2$ and the threshold parameter
$s_0$. There lack experimental data for the phenomenological
hadronic spectral densities of the tetraquark states, the present
predictions can't be confronted with the experimental data.

In
Refs.\cite{Maiani2004,Maiani20042,Maiani2005,Maiani2008,Polosa0902},
Maiani et al take the diquarks as the basic constituents, examine
the rich spectrum of the
 diquark-antidiquark states  from the constituent diquark masses and the spin-spin
 interactions, and try to  accommodate some of the newly observed charmonium-like resonances not
 fitting a pure $c\bar{c}$ assignment. The predictions depend heavily on  the assumption that the light
 scalar mesons $a_0(980)$ and $f_0(980)$ are tetraquark states,
 the  basic  parameters (constituent diquark masses) are
 estimated thereafter.
 In the conventional quark models, the
constituent quark masses  are taken as the basic input parameters,
and fitted to reproduce the mass spectra  of the conventional mesons
and baryons. However, the present experimental knowledge about the
phenomenological hadronic spectral densities of the multiquark
states is  rather vague, even existence of the multiquark states is
not confirmed with confidence, and no knowledge about either there
are high resonances or not. The predicted constituent diquark masses
can not be confronted with the experimental data.

The LHCb is a dedicated $b$ and $c$-physics precision experiment at
the LHC (large hadron collider). The LHC will be the world's most
copious  source of the $b$ hadrons, and  a complete spectrum of the
$b$ hadrons will be available through gluon fusion. In proton-proton
collisions at $\sqrt{s}=14\,\rm{TeV}$¡Ì, the $b\bar{b}$ cross
section is expected to be $\sim 500\mu b$ producing $10^{12}$
$b\bar{b}$ pairs in a standard  year of running at the LHCb
operational luminosity of $2\times10^{32} \rm{cm}^{-2}
\rm{sec}^{-1}$ \cite{LHC}. The vector tetraquark states predicted in
the present work may be observed at the LHCb, if they exist  indeed.
We can search for the vector hidden charm tetraquark states  in the
$D\bar{D}$, $D\bar{D^*}$, $D^*\bar{D^*}$, $D_s\bar{D_s}$,
$D_s\bar{D^*_s}$, $D_s^*\bar{D_s^*}$, $J/\psi \rho$, $J/\psi \phi$,
$J/\psi \omega$, $J/\psi \pi$, $J/\psi f_0(980)$, $J/\psi K$,
$\eta_c\pi$, $\eta_c\eta$, $\cdots$ invariant mass distributions and
search for the vector hidden bottom tetraquark states in the
$B\bar{B}$, $B\bar{B^*}$, $B^*\bar{B^*}$, $B_s\bar{B_s}$,
$B_s\bar{B_s^*}$, $B_s^*\bar{B_s^*}$, $\Upsilon \rho$, $\Upsilon
\phi$, $\Upsilon \omega$, $\Upsilon \pi$, $\Upsilon K$, $\Upsilon
f_0(980)$, $\eta_b\pi$, $\eta_b\eta$, $\cdots$ invariant mass
distributions.

\section{Conclusion}
In this article, we study the mass spectrum of the vector hidden
charm and bottom tetraquark states  with the QCD sum rules.  The
mass spectrum are  calculated  by imposing the two criteria (pole
dominance and convergence of the operator product expansion) of the
QCD sum rules.  As there lack experimental data for the
phenomenological hadronic spectral densities of the tetraquark
states, the present predictions can't be confronted with the
experimental data. We can search for the vector  hidden charm and
bottom tetraquark states at the LHCb  or the Fermi-lab Tevatron.

\section*{Appendix}
The spectral densities at the level of the quark-gluon degrees of
freedom:

\begin{eqnarray}
\rho^{\pm}_{q\bar{q}}(s)&=&\frac{1}{3072 \pi^6}
\int_{\alpha_{min}}^{\alpha_{max}}d\alpha
\int_{\beta_{min}}^{1-\alpha} d\beta
\alpha\beta(1-\alpha-\beta)^3(s-\widetilde{m}^2_Q)^2(35s^2-26s\widetilde{m}^2_Q+3\widetilde{m}^4_Q)
\nonumber \\
&&\pm\frac{ m_Q\langle \bar{q}q\rangle}{32 \pi^4}
\int_{\alpha_{min}}^{\alpha_{max}}d\alpha
\int_{\beta_{min}}^{1-\alpha} d\beta
(1-\alpha-\beta)(s-\widetilde{m}^2_Q) \left[(4\beta-3\alpha) s+(\alpha-2\beta)\widetilde{m}^2_Q\right] \nonumber\\
&& \pm\frac{ m_Q\langle \bar{q}g_s\sigma Gq\rangle}{64 \pi^4}
\int_{\alpha_{min}}^{\alpha_{max}}d\alpha
\int_{\beta_{min}}^{1-\alpha} d\beta
\left[(2\alpha-3\beta)s-(\alpha-2\beta)\widetilde{m}^2_Q) \right] \nonumber\\
&&-\frac{m_Q^2\langle \bar{q}q\rangle^2}{12 \pi^2}
\int_{\alpha_{mix}}^{\alpha_{max}} d\alpha
+\frac{m_Q^2\langle\bar{q}q\rangle\langle\bar{q}g_s \sigma
Gq\rangle}{24\pi^2}\int_{\alpha_{mix}}^{\alpha_{max}} d\alpha \left[
1+\frac{s}{M^2}\right]\delta(s-\widetilde{\widetilde{m}}_Q^2)
\nonumber \\
&&-\frac{m_Q^2\langle\bar{q}g_s \sigma
Gq\rangle^2}{192\pi^2M^6}\int_{\alpha_{mix}}^{\alpha_{max}} d\alpha
\widetilde{\widetilde{m}}_Q^4\delta(s-\widetilde{\widetilde{m}}_Q^2)\,
,
\end{eqnarray}

\begin{eqnarray}
\rho^{\pm}_{q\bar{s}}(s)&=&\frac{1}{3072 \pi^6}
\int_{\alpha_{min}}^{\alpha_{max}}d\alpha
\int_{\beta_{min}}^{1-\alpha} d\beta
\alpha\beta(1-\alpha-\beta)^3(s-\widetilde{m}^2_Q)^2(35s^2-26s\widetilde{m}^2_Q+3\widetilde{m}^4_Q)
\nonumber \\
&&\mp\frac{ m_sm_Q}{256 \pi^6}
\int_{\alpha_{min}}^{\alpha_{max}}d\alpha
\int_{\beta_{min}}^{1-\alpha} d\beta \beta
(1-\alpha-\beta)^2(s-\widetilde{m}^2_Q)^2(5s-2\widetilde{m}^2_Q)   \nonumber\\
&&+\frac{ m_s\langle \bar{s}s\rangle}{64 \pi^4}
\int_{\alpha_{min}}^{\alpha_{max}}d\alpha
\int_{\beta_{min}}^{1-\alpha} d\beta \alpha \beta
(1-\alpha-\beta)(15s^2-16s\widetilde{m}^2_Q+3\widetilde{m}^4_Q)   \nonumber\\
&&\pm\frac{ m_Q\langle \bar{q}q\rangle}{32 \pi^4}
\int_{\alpha_{min}}^{\alpha_{max}}d\alpha
\int_{\beta_{min}}^{1-\alpha} d\beta \alpha
(1-\alpha-\beta)(s-\widetilde{m}^2_Q) (\widetilde{m}^2_Q-3s) \nonumber\\
&&\mp\frac{ m_Q\langle \bar{s}s\rangle}{16 \pi^4}
\int_{\alpha_{min}}^{\alpha_{max}}d\alpha
\int_{\beta_{min}}^{1-\alpha} d\beta \beta
(1-\alpha-\beta) (s-\widetilde{m}^2_Q) (\widetilde{m}^2_Q-2s) \nonumber\\
&& \pm\frac{ m_Q\langle \bar{q}g_s\sigma Gq\rangle}{64 \pi^4}
\int_{\alpha_{min}}^{\alpha_{max}}d\alpha
\int_{\beta_{min}}^{1-\alpha} d\beta \alpha
(2s-\widetilde{m}^2_Q)  \nonumber\\
&& \mp \frac{ m_Q\langle \bar{s}g_s\sigma Gs\rangle}{64 \pi^4}
\int_{\alpha_{min}}^{\alpha_{max}}d\alpha
\int_{\beta_{min}}^{1-\alpha} d\beta \beta
(3s-2\widetilde{m}^2_Q)  \nonumber\\
&&-\frac{ m_s\langle \bar{s}g_s\sigma Gs\rangle}{192 \pi^4}
\int_{\alpha_{min}}^{\alpha_{max}}d\alpha
\int_{\beta_{min}}^{1-\alpha} d\beta \alpha \beta
\left[8s-3\widetilde{m}^2_Q+s^2\delta(s-\widetilde{m}^2_Q)\right]   \nonumber\\
&& +\frac{ m_sm_Q^2\langle \bar{q}q\rangle}{16 \pi^4}
\int_{\alpha_{min}}^{\alpha_{max}}d\alpha
\int_{\beta_{min}}^{1-\alpha} d\beta
(s-\widetilde{m}^2_Q)  \nonumber\\
&&-\frac{m_Q^2\langle \bar{q}q\rangle \langle \bar{s}s\rangle}{12
\pi^2} \int_{\alpha_{mix}}^{\alpha_{max}} d\alpha
-\frac{m_sm_Q^2\langle \bar{q}g_s\sigma Gq\rangle }{64 \pi^4}
\int_{\alpha_{mix}}^{\alpha_{max}} d\alpha \nonumber\\
&&\mp\frac{m_sm_Q\langle \bar{q}q\rangle \langle \bar{s}s\rangle}{24
\pi^2} \int_{\alpha_{mix}}^{\alpha_{max}} d\alpha \alpha
\left[2+s\delta(s-\widetilde{\widetilde{m}}^2_Q) \right]\nonumber\\
&&+\frac{m_Q^2\left[\langle\bar{q}q\rangle\langle\bar{s}g_s \sigma
Gs\rangle+\langle\bar{s}s\rangle\langle\bar{q}g_s \sigma
Gq\rangle\right]}{48\pi^2}\int_{\alpha_{mix}}^{\alpha_{max}} d\alpha
\left[ 1+\frac{s}{M^2}\right]\delta(s-\widetilde{\widetilde{m}}_Q^2)
\nonumber \\
&&\pm\frac{m_sm_Q\left[2\langle\bar{q}q\rangle\langle\bar{s}g_s
\sigma Gs\rangle+3\langle\bar{s}s\rangle\langle\bar{q}g_s \sigma
Gq\rangle\right]}{288\pi^2M^2}\int_{\alpha_{mix}}^{\alpha_{max}}
d\alpha \alpha\left[
s-\frac{s^2}{M^2}\right]\delta(s-\widetilde{\widetilde{m}}_Q^2)
\nonumber \\
&&-\frac{m_Q^2\langle\bar{q}g_s \sigma Gq\rangle\langle\bar{s}g_s
\sigma Gs\rangle}{192\pi^2M^6}\int_{\alpha_{mix}}^{\alpha_{max}}
d\alpha
\widetilde{\widetilde{m}}_Q^4\delta(s-\widetilde{\widetilde{m}}_Q^2)
\, ,
\end{eqnarray}

\begin{eqnarray}
\rho^{\pm}_{s\bar{s}}(s)&=&\frac{1}{3072 \pi^6}
\int_{\alpha_{min}}^{\alpha_{max}}d\alpha
\int_{\beta_{min}}^{1-\alpha} d\beta
\alpha\beta(1-\alpha-\beta)^3(s-\widetilde{m}^2_Q)^2(35s^2-26s\widetilde{m}^2_Q+3\widetilde{m}^4_Q)
\nonumber \\
&&\pm \frac{ m_sm_Q}{256 \pi^6}
\int_{\alpha_{min}}^{\alpha_{max}}d\alpha
\int_{\beta_{min}}^{1-\alpha} d\beta
(1-\alpha-\beta)^2(s-\widetilde{m}^2_Q)^2\left[(4\alpha-5\beta)s-(\alpha-2\beta)\widetilde{m}^2_Q \right]   \nonumber\\
&&+\frac{ m_s\langle \bar{s}s\rangle}{32 \pi^4}
\int_{\alpha_{min}}^{\alpha_{max}}d\alpha
\int_{\beta_{min}}^{1-\alpha} d\beta \alpha \beta
(1-\alpha-\beta)(15s^2-16s\widetilde{m}^2_Q+3\widetilde{m}^4_Q)   \nonumber\\
&&\pm\frac{ m_Q\langle \bar{s}s\rangle}{16 \pi^4}
\int_{\alpha_{min}}^{\alpha_{max}}d\alpha
\int_{\beta_{min}}^{1-\alpha} d\beta
(1-\alpha-\beta) (s-\widetilde{m}^2_Q) \left[(4\beta-3\alpha)s+(\alpha-2\beta)\widetilde{m}^2_Q \right]\nonumber\\
&& \pm\frac{ m_Q\langle \bar{s}g_s\sigma Gs\rangle}{64 \pi^4}
\int_{\alpha_{min}}^{\alpha_{max}}d\alpha
\int_{\beta_{min}}^{1-\alpha} d\beta
\left[(2\alpha-3\beta)s-(\alpha-2\beta)\widetilde{m}^2_Q \right]  \nonumber\\
&&-\frac{ m_s\langle \bar{s}g_s\sigma Gs\rangle}{96 \pi^4}
\int_{\alpha_{min}}^{\alpha_{max}}d\alpha
\int_{\beta_{min}}^{1-\alpha} d\beta \alpha \beta
\left[8s-3\widetilde{m}^2_Q+s^2\delta(s-\widetilde{m}^2_Q)\right]   \nonumber\\
&& +\frac{ m_sm_Q^2\langle \bar{s}s\rangle}{8 \pi^4}
\int_{\alpha_{min}}^{\alpha_{max}}d\alpha
\int_{\beta_{min}}^{1-\alpha} d\beta
(s-\widetilde{m}^2_Q)  \nonumber\\
&&-\frac{m_Q^2  \langle \bar{s}s\rangle^2}{12 \pi^2}
\int_{\alpha_{mix}}^{\alpha_{max}} d\alpha -\frac{m_sm_Q^2\langle
\bar{s}g_s\sigma Gs\rangle }{32 \pi^4}
\int_{\alpha_{mix}}^{\alpha_{max}} d\alpha \nonumber\\
&&+\frac{m_Q^2\langle\bar{s}s\rangle\langle\bar{s}g_s \sigma
Gs\rangle}{24\pi^2}\int_{\alpha_{mix}}^{\alpha_{max}} d\alpha \left[
1+\frac{s}{M^2}\right]\delta(s-\widetilde{\widetilde{m}}_Q^2)\nonumber\\
&&\pm\frac{5m_sm_Q\langle\bar{s}s\rangle\langle\bar{s}g_s \sigma
Gs\rangle}{288\pi^2M^2}\int_{\alpha_{mix}}^{\alpha_{max}} d\alpha
\alpha \left[
s-\frac{s^2}{M^2}\right]\delta(s-\widetilde{\widetilde{m}}_Q^2)
\nonumber\\
&&\mp\frac{5m_sm_Q\langle\bar{s}s\rangle\langle\bar{s}g_s \sigma
Gs\rangle}{144\pi^2}\int_{\alpha_{mix}}^{\alpha_{max}} d\alpha
(1-\alpha) \left[1+
\frac{s}{M^2}+\frac{s^2}{2M^4}\right]\delta(s-\widetilde{\widetilde{m}}_Q^2)
\nonumber \\
&&-\frac{m_Q^2\langle\bar{s}g_s \sigma
Gs\rangle^2}{192\pi^2M^6}\int_{\alpha_{mix}}^{\alpha_{max}} d\alpha
\widetilde{\widetilde{m}}_Q^4\delta(s-\widetilde{\widetilde{m}}_Q^2)\,
,
\end{eqnarray}

\section*{Acknowledgements}
This  work is supported by National Natural Science Foundation of
China, Grant Number 10775051, and Program for New Century Excellent
Talents in University, Grant Number NCET-07-0282.


\begin{thebibliography}{99}


\bibitem{review1} E. S. Swanson, Phys. Rept. {\bf 429} (2006) 243.

\bibitem{review2} E. Klempt and A. Zaitsev, Phys. Rept. {\bf 454} (2007) 1.

\bibitem{review3} M. B. Voloshin, Prog. Part. Nucl. Phys. {\bf 61} (2008) 455.


\bibitem{review4} S. Godfrey and  S. L. Olsen, Ann. Rev. Nucl. Part. Sci. {\bf 58} (2008) 51.

\bibitem{Olsen2009}  S. L. Olsen,  arXiv:0901.2371.

\bibitem{Belle-z4430} S. K. Choi et al, Phys. Rev. Lett. {\bf 100} (2008) 142001.

\bibitem{Babar0811} B. Aubert et al, arXiv:0811.0564.

\bibitem{Belle-chipi}  R. Mizuk  et al,  Phys. Rev. {\bf D78} (2008) 072004.


\bibitem{Wang0807} Z. G. Wang, Eur. Phys. J. {\bf C59} (2009) 675.

\bibitem{Wang08072} Z. G. Wang, arXiv:0807.4592.

\bibitem{Wang0902} Z. G. Wang, arXiv:0902.2062.


\bibitem{SVZ79}  M. A. Shifman, A. I. Vainshtein and V. I. Zakharov,
Nucl. Phys. {\bf B147} (1979) 385, 448.

\bibitem{Reinders85} L. J. Reinders, H. Rubinstein and S. Yazaki, Phys. Rept. {\bf 127} (1985) 1.


\bibitem{LHC}  G. Kane and A. Pierce, "Perspectives On LHC Physics",
World Scientific Publishing Company,  2008.

\bibitem{Jaffe2003} R. L. Jaffe and  F. Wilczek, Phys. Rev. Lett. {\bf 91} (2003) 232003.


\bibitem{Jaffe2004} R. L. Jaffe, Phys. Rept. {\bf 409} (2005) 1.

\bibitem{GI1} A. De Rujula, H. Georgi and S. L. Glashow, Phys. Rev.  {\bf D12}
(1975) 147.

\bibitem{GI2} T. DeGrand, R. L. Jaffe, K. Johnson and J. E. Kiskis,
Phys.  Rev.  {\bf D12} (1975) 2060.


\bibitem{Wang1} Z. G. Wang, Nucl. Phys. {\bf A791} (2007) 106.


\bibitem{Wang2} Z. G. Wang, W. M. Yang and S. L. Wan, J. Phys. {\bf G31} (2005) 971.

\bibitem{Wang0904} Z. G. Wang, arXiv:0903.5200.

\bibitem{Hbaryon} N. Kodama, M. Oka and T. Hatsuda, Nucl. Phys. {\bf A580} (1994) 445.


\bibitem{Ioffe2005} B. L. Ioffe, Prog. Part. Nucl. Phys. {\bf 56} (2006)
232.

\bibitem{Kho9801} A. Khodjamirian and R. Ruckl, Adv. Ser. Direct. High Energy Phys. {\bf 15} (1998) 345.
\bibitem{PDG} C. Amsler et al, Phys. Lett. {\bf  B667} (2008) 1.

\bibitem{Wang0708} Z. G. Wang, Chin. Phys. {\bf C32} (2008) 797.


\bibitem{Maiani2004} L. Maiani, F. Piccinini, A. D. Polosa and V. Riquer, Phys. Rev. Lett. {\bf 93} (2004) 212002.

\bibitem{Maiani20042} L. Maiani, F. Piccinini, A. D. Polosa and V. Riquer, Phys. Rev. {\bf D71} (2005) 014028.

\bibitem{Maiani2005} L. Maiani, F. Piccinini, A. D. Polosa and V. Riquer, Phys. Rev. {\bf D72} (2005) 031502.

\bibitem{Maiani2008} L. Maiani, A. D. Polosa and V. Riquer, New J. Phys. {\bf 10} (2008) 073004.

\bibitem{Polosa0902}  N. V. Drenska, R. Faccini and  A. D. Polosa, arXiv:0902.2803.


\end{thebibliography}
\end{document}